\documentclass[11pt,a4paper]{article} 
\pdfoutput=1
\usepackage{jheppub} 
\usepackage{graphicx} 
\usepackage{amsmath} 
\usepackage{wasysym} 
\usepackage{amssymb} 
\usepackage{slashed} 
\usepackage{bigstrut} 
\usepackage{mathtools} 
\usepackage{multirow} 
\usepackage[utf8]{inputenc}
\usepackage{lineno}
\usepackage{todonotes}
\usepackage{xspace}
\usepackage{lscape}
\usepackage[calc,useregional]{datetime2}

\newcommand{\chisq}{\ensuremath{\chi^{2}}\xspace}
\newcommand{\afb}{\ensuremath{A_{\text{FB}}}\xspace}
\newcommand{\new}[1]{{#1}}
\newcommand{\neww}[1]{{#1}}
\newcommand{\newepjc}[1]{{#1}}

\title{

  Exploring SMEFT Couplings Using the Forward-Backward Asymmetry in Neutral Current Drell-Yan Production at the LHC
}

\author[a,b]{Andrii Anataichuk,}
\author[c]{Sven-Olaf Moch,}
\author[d,e]{and the xFitter Developers' team: Hamed Abdolmaleki,}
\author[f]{Simone Amoroso,}
\author[g]{Daniel Britzger,}
\author[f]{Filippo Dattola,}
\author[h]{Juri Fiaschi,}
\author[i]{Francesco Giuli,}
\author[f]{Alexander Glazov,}
\author[i,j,k]{Francesco Hautmann,}
\author[l]{Agnieszka Luszczak,}
\author[f]{Sara Taheri Monfared,}
\author[m]{Fred Olness,}
\author[f]{Federico Vazzoler,}
\author[c]{Oleksandr Zenaiev}

\affiliation[a]{V.~N. Karazin Kharkiv National University, Kharkiv, Ukraine (on leave)}
\affiliation[b]{Taras Shevchenko National University of Kyiv, Kyiv, Ukraine}
\affiliation[c]{Hamburg University, Hamburg, Germany}
\affiliation[d]{School of Physics, Institute for Research in Fundamental Sciences (IPM), Tehran, Iran}
\affiliation[e]{Department of Physics, Faculty of Science, Malayer University, Malayer, Iran}
\affiliation[f]{Deutsches Elektronen-Synchrotron DESY, Germany}
\affiliation[g]{Max Planck Institute, Munich, Germany}
\affiliation[h]{Universit\`{a} degli Studi di Milano-Bicocca \& INFN, Milano, Italy}
\affiliation[i]{CERN, Geneva, Switzerland}
\affiliation[j]{University of Antwerp, Antwerpen, Belgium}
\affiliation[k]{University of Oxford, Oxford, UK}
\affiliation[l]{Cracow University of Technology, Cracow, Poland}
\affiliation[m]{SMU Physics, Dallas, USA}

\emailAdd{andanataychuk@gmail.com} 
\emailAdd{sven-olaf.moch@desy.de} 
\emailAdd{Hamed.Abdolmaleki@desy.de}
\emailAdd{simone.amoroso@desy.de} 
\emailAdd{britzger@mpp.mpg.de} 
\emailAdd{filippo.dattola@desy.de} 
\emailAdd{Juri.Fiaschi@liverpool.ac.uk}
\emailAdd{francesco.giuli@cern.ch}
\emailAdd{alexander.glazov@desy.de}
\emailAdd{hautmann@thphys.ox.ac.uk}
\emailAdd{agnieszka.luszczak@desy.de}
\emailAdd{taheri@mail.desy.de}
\emailAdd{olness@smu.edu}
\emailAdd{federico.vazzoler@cern.ch}
\emailAdd{oleksandr.zenaiev@desy.de}

\abstract{
  Neutral current Drell-Yan (DY) lepton-pair production is considered to study $Z$-boson quark couplings.
	Using the open-source fit platform   \texttt{xFitter},  
	we  investigate the impact of high-statistics 
	measurements of the neutral current DY (NCDY) forward-backward asymmetry 
	$A_{\rm{FB}}$ 
	near the weak boson mass scale 
	in the present and forthcoming stages of the 
	Large Hadron Collider (LHC).  Besides recovering earlier results on 
	the $A_{\rm{FB}}$ sensitivity to parton distribution functions,  
	we analyze the precision determination of  $Z$-boson couplings to 
	left-handed and right-handed $u$-quarks and $d$-quarks, and explore  
	Beyond-Standard-Model contributions using the Standard Model Effective Field Theory (SMEFT) framework. 
	We perform a sensitivity study and
	comment on the role of the $A_{\rm{FB}}$ asymmetry  for 
	the electroweak SMEFT fit  and precision $Z$-boson physics 
	at the LHC and  high-luminosity HL-LHC.
}


\preprint{DESY-23-160}
\arxivnumber{arXiv:2310.19638} 
\begin{document} 
\maketitle

\section{Introduction}
\label{intro}

The  physics program centering 
on  electroweak (EW) precision observables receives 
essential inputs from  measurements of $W$ 
and $Z$ bosons at the LHC.  
Owing  to the cancellation of 
many systematic uncertainties, 
the  forward-backward asymmetry $A_{\rm{FB}}$ 
in NCDY lepton-pair production 
is a crucial component of this program. The $A_{\rm{FB}}$ asymmetry 
is employed for  
 determinations of the weak mixing angle $\theta_W$
from LHC 
measurements at the $Z$-pole~\cite{CMS:2018ktx,CMS:2022uul,ATLAS:2018gqq,ATLAS:2018qvs},  
complementing LEP/SLD~\cite{ALEPH:2005ab}  
and Tevatron~\cite{CDF:2018cnj} results.  

Given that  parton distribution functions (PDFs) constitute one of the 
dominant uncertainty sources in the 
precision EW physics program at the LHC, 
 it is especially relevant that the $A_{\rm{FB}}$ asymmetry  has 
 been shown  to provide us with new sensitivity to 
 PDFs~\cite{Accomando:2017scx,Accomando:2018nig,Accomando:2019vqt,Abdolmaleki:2019ubu,Fiaschi:2021okg}.  
 This sensitivity is currently not exploited in global PDF 
  extractions~\cite{Hou:2019efy,Bailey:2020ooq,NNPDF:2021njg,Alekhin:2017kpj,H1:2015ubc},  
  and could potentially lead to 
  dramatic improvements in our knowledge of PDFs. 
  This applies, 
  in particular, in kinematic regions which are  
  relevant for new physics searches in the 
multi-TeV region at the LHC, for instance 
in the context of Beyond-Standard-Model (BSM) heavy 
$Z^\prime$~\cite{Fiaschi:2022wgl,Accomando:2019ahs} and 
$W^\prime$ bosons~\cite{Fiaschi:2021sin}, and   
photon-induced di-lepton 
production 
processes~\cite{Accomando:2016ehi,Accomando:2016ouw,Accomando:2016tah}.  

Furthermore, as in Ref.~\cite{Fiaschi:2021okg}  the impact of the 
forward-backward $A_{\rm{FB}}$ asymmetry in the neutral current sector 
may be combined with that of the lepton-charge $A_W$ asymmetry 
in the charged current sector.  This points to 
strategies which are alternative to those taken in experimental analyses 
such as in Refs.~\cite{CMS:2018ktx,ATLAS:2018gqq,Bodek:2015ljm},  
and  aim at exploiting new measurements, capable of 
providing  sensitivity to PDFs with low 
theoretical  and experimental systematics while controlling correlations. 
Related investigations of the $A_{\rm{FB}}$ asymmetry 
in Ref.~\cite{Ball:2022qtp} focus on the behaviour induced by the 
NNPDF4.0 set~\cite{NNPDF:2021njg}.   
See also the 
studies~\cite{Fu:2023rrs,Yang:2022zvx,Yang:2022bxv,Xie:2022tzo,Yang:2021cpd,Fu:2020mxl} 
based on the package {\sc ePump}~\cite{Schmidt:2018hvu,Willis:2018yln}.    

To systematically investigate the role of the
asymmetry in precision EW measurements,
searches for BSM phenomena, and determinations of PDFs, a well-established framework is provided by the
Standard Model Effective Field Theory (SMEFT)~\cite{Buchmuller:1985jz}. Details can be found in recent reviews~\cite{Isidori:2023pyp,Brivio:2017vri}
and SMEFT fitting packages~\cite{Ethier:2021bye,Ellis:2020unq,Brivio:2017btx}. 
Recent SMEFT studies of precision electroweak 
observables in di-lepton channels at the LHC have been performed in 
Refs.~\cite{Bellafronte:2023amz,Dawson:2018dxp,Boughezal:2023nhe,Boughezal:2021tih,Breso-Pla:2021qoe,Allwicher:2022gkm} and 
analogous studies on the role of PDFs in BSM searches in Refs.~\cite{Greljo:2021kvv,Hammou:2023heg}. 
 
In this paper we will concentrate 
on $A_{\rm{FB}}$ asymmetry measurements 
in NCDY production 
in the region near the $Z$-boson mass scale. The analysis will be performed in the framework of the SMEFT 
Lagrangian, including operators up to dimension 
$D = 6$~\cite{Buchmuller:1985jz,Grzadkowski:2010es}, 
\begin{equation} 
\label{dim6lagr} 
    {\cal L}
    =    {\cal L}^{\rm{(SM)}} + 
    {1 \over \Lambda^2}   \sum_{j=1}^{N_6} C_j^{\rm{(6)}} \ {\cal O}_j^{\rm{(6)}}    \; ,   
\end{equation} 
where the first term on the right hand side is the SM Lagrangian, consisting of 
operators of mass dimension $D=4$, while 
the next term is the EFT contribution  containing  
$N_6$ operators $  {\cal O}_j$ of mass dimension $D = 6$, each 
weighted by the dimensionless Wilson coefficient $C_j$ divided by 
$\Lambda^2$, where  $\Lambda$ is  the  ultraviolet mass scale of the EFT. 
 
In the di-lepton mass region near the  $Z$-boson  
peak, four-fermion operators and 
dipole operators coupling fermions and vector bosons 
can be neglected~\cite{Bellafronte:2023amz,Breso-Pla:2021qoe} in 
Eq.~(\ref{dim6lagr}), and 
the whole  effect of the $D = 6$ SMEFT Lagrangian is a modification of the 
vector boson couplings to fermions. Using 
LEP constraints~\cite{ALEPH:2005ab},  
corrections to $Z$-boson couplings to 
leptons can also be 
neglected~\cite{Breso-Pla:2021qoe}. 
We will thus focus  on the 
SMEFT corrections to  $Z$-boson couplings to 
$u$-type (including $c$)  and $d$-type (including $s$, $b$) quarks, 
that are least constrained by LEP and have not comprehensively been studied at the LHC. 

To explore these SMEFT couplings, we will extend the 
implementation  of 
the $A_{\rm{FB}}$ asymmetry 
provided  in Ref.~\cite{Accomando:2019vqt}, using   
the quantum chromodynamics (QCD) fit platform    \texttt{xFitter} (formerly known as  \texttt{HERAFitter})~\cite{Alekhin:2014irh,xFitter:2022zjb,xfittergitlab}.   
As a check, we will recover the results of Ref.~\cite{Accomando:2019vqt} on  PDF extracted 
from $A_{\rm{FB}}$ pseudodata, and in addition we will obtain new constraints on 
$Z$-boson vector and axial couplings. We will 
 examine the projected luminosity scenario
 of 3000 fb$^{-1}$ for the High-Luminosity LHC (HL-LHC)~\cite{Gianotti:2002xx,Azzi:2019yne,CidVidal:2018eel}.%, 

The paper is organized as follows. In Sec.~\ref{sec:xfitter} we present  the SMEFT 
treatment of the NCDY di-lepton production process in terms of 
EFT corrections to the SM couplings, and its implementation to make 
predictions for the $A_{\rm{FB}}$ asymmetry  in    \texttt{xFitter}. 
In Sec.~\ref{sec:pseudodata} we describe the  $A_{\rm{FB}}$ pseudodata generation. 
In Sec.~\ref{sec:results} we carry out the main   analysis within  \texttt{xFitter}, leading  
to the determination of the SMEFT couplings.  
In Sec.~\ref{sec:conclusions} we give conclusions. 

\section{$A_{\rm{FB}}$ within SMEFT in {\tt xFitter}}
\label{sec:xfitter} 

In this section we start by describing 
 the $D = 6$ SMEFT Lagrangian for 
$Z$-boson interactions with fermions, and introduce the SMEFT 
couplings for left-handed and right-handed $u$-quarks and $d$-quarks. 
Next we define the SMEFT vector and axial couplings, and express 
the forward-backward asymmetry $A_{\rm{FB}}$ in terms of these 
couplings.  We discuss the extension of the  \texttt{xFitter} 
implementation~\cite{Accomando:2019vqt} for  $A_{\rm{FB}}$ 
to the SMEFT case.

The SMEFT Lagrangian for the coupling of the $Z$-boson to fermions is given by 
\begin{eqnarray} 
\label{smeftL_Z} 
{\cal L}_Z^{\rm{(SMEFT)}} &=& 2 M_Z \sqrt{ G_\mu \sqrt{2} } 
\  Z^\alpha   
\left\{ 
{\overline{q}}_L \gamma_\alpha 
\left( 
g_{L ({\rm{SM}})}^{Zq} +  \delta g_L^{Zq}  
\right)  q_L  
+ {\overline{u}}_R \gamma_\alpha \left( g_{R ({\rm{SM}})}^{Zu}    
+ \delta g_R^{Zu}  \right)  u_R \right. 
\nonumber\\ 
&+&  \left. 
 {\overline{d}}_R \gamma_\alpha 
\left( g_{R ({\rm{SM}})}^{Zd} +  \delta g_R^{Zd}  
\right)  d_R 
+ \{ {\rm{leptonic}} \,  {\rm{terms}}  \} 
\right\} 
\end{eqnarray} 
Here $q_L$ is the left-handed quark SU(2) doublet, while 
$u_R$ and $d_R$ are the right-handed quark SU(2) singlets. 
The left-handed and right-handed quark 
SM couplings   are expressed in terms of the 
weak mixing angle $\theta_W$ as follows, 
\begin{eqnarray} 
\label{gLR_SM}
&&  g_{R ({\rm{SM}})}^{Zu} = 1/2 - 2/3 \sin^2 \theta_W 
\; , \;\;\;\;   g_{L ({\rm{SM}})}^{Zu} = - 2/3 \sin^2 \theta_W \; , 
\nonumber\\ 
&&  g_{R ({\rm{SM}})}^{Zd} = -1/2 + 1/3 \sin^2 \theta_W 
\; , \;\;\;\;   g_{L ({\rm{SM}})}^{Zd} =  1/3 \sin^2 \theta_W  \; .   
\end{eqnarray}  
The SMEFT couplings are obtained from the SM couplings via 
the corrections $\delta g$, i.e., 
$g_{ ({\rm{SMEFT}})} \equiv  g_{ ({\rm{SM}})} + \delta g$:  
\begin{eqnarray} 
\label{gLR_SMEFT}
&&    g_L^{Zu}  \equiv  g_{L ({\rm{SMEFT}})}^{Zu}
=  g_{L ({\rm{SM}})}^{Zu} + \delta g_L^{Zu}
\; , \;\;\;\;  g_R^{Zu} \equiv    g_{R ({\rm{SMEFT}})}^{Zu}  =  
g_{R ({\rm{SM}})}^{Zu}
+ \delta g_R^{Zu}   \; , 
\nonumber\\ 
&&   g_L^{Zd}  \equiv  g_{L ({\rm{SMEFT}})}^{Zd}   = 
g_{L ({\rm{SM}})}^{Zd}
+ \delta g_L^{Zd}
\; , \;\;\;\;    g_R^{Zd}  \equiv  g_{R ({\rm{SMEFT}})}^{Zd} = 
g_{R ({\rm{SM}})}^{Zd} 
+ \delta g_R^{Zd} \; . 
\end{eqnarray}
\neww{In our analysis, we assume $\delta g_{R,L}^{Zd} = \delta g_{R,L}^{Zs} = \delta g_{R,L}^{Zb}$ and $\delta g_{R,L}^{Zu} = \delta g_{R,L}^{Zc}$.
The contributions from heavy $c$- and $b$-quarks are small, of the order of 10\%.}
The vector and axial couplings of the $Z$-boson are defined by taking the 
combinations $ L \pm R$ of the left-handed and right-handed fermion couplings. 
So the SMEFT vector and axial 
couplings are given by 
\begin{eqnarray} 
\label{gVA_SMEFT}
&&  g_{V }^{Zu} =  g_{R }^{Zu} + g_{L }^{Zu}
\; , \;\;\;\;  g_{A }^{Zu} =    
g_{R }^{Zu} - g_{L }^{Zu}  \; , 
\nonumber\\ 
&&  g_{V }^{Zd} =  g_{R }^{Zd} + g_{L }^{Zd}
\; , \;\;\;\;  g_{A }^{Zd} =    
g_{R }^{Zd} - g_{L }^{Zd} \; . 
\end{eqnarray}

In order to maximize the sensitivity, we consider  the 
 DY triple-differential cross section in the  di-lepton invariant mass 
$M_{\ell\ell}$, di-lepton rapidity $y_{\ell\ell}$  and angular variable 
$\theta^{*}$ between the outgoing lepton and the incoming quark in 
the Collins-Soper (CS) reference frame~\cite{Collins:1977iv}. 
In this frame, the decay angle is measured from an axis symmetric with respect to the two incoming partons. The 
expression for the angle $\theta^{*}$ in the CS frame is given by 
\begin{equation}
\cos\theta^{*} = \dfrac{p_{{Z},\ell\ell}}{M_{\ell\ell}|p_{{Z},\ell\ell}|} \dfrac{p_{1}^{+}p_{2}^{-}-p_{1}^{-}p_{2}^{+}}{\sqrt{M^{2}_{\ell\ell}+p^{2}_{{T},\ell\ell}}},
\end{equation}
where $p_{i}^{\pm}=E_{i}\pm p_{{Z},i}$ and the index $i = 1,2$ corresponds to the positive and negative charged lepton respectively. Here $E$ and $p_{{Z}}$ are the energy and the $z$-components of the leptonic four-momentum, respectively; $p_{{Z},\ell\ell}$ is the di-lepton $z$-component of the momentum and $p_{{T},\ell\ell}$ is the di-lepton transverse momentum. 
At leading order (LO) in QCD and EW theory, 
this cross section can be written as 
\begin{equation}
 \label{eq:triple_diff}
 \frac{d^3 \sigma}{dM_{\ell\ell}dy_{\ell\ell}d\cos\theta^*} = \frac{\pi\alpha^2}{3M_{\ell\ell}s} \sum_q P_q \left[f_q (x_1, Q^2) f_{\bar{q}} (x_2, Q^2) + f_{\bar{q}} (x_1, Q^2) f_q (x_2, Q^2)\right],
\end{equation}
%\noindent
where $s$ is the square of the centre-of-mass energy of the collision,  
 $x_{1,2} = M_{\ell\ell} e^{\pm y_{\ell\ell}}/\sqrt{s}$ are the  momentum 
 fractions  of the initial-state partons, 
 $f_{q,\bar{q}} (x_i, Q^2)$ are their PDFs, $Q^2$ is the squared factorization scale, and 
the factor $P_q$ contains the propagators and couplings of the $Z$-boson, photon, and $Z$-$\gamma$ interference, 
\begin{align}
 \label{eq:propagators}
 P_q &= e^2_\ell e^2_q (1 + \cos^2\theta^*)
 \\ 
 \nonumber
 &+ \frac{M^2_{\ell\ell}(M^2_{\ell\ell} - M^2_Z)}{2 \sin^2\theta_W \cos^2\theta_W\left[(M^2_{\ell\ell} - M^2_Z)^2 + \Gamma^2_Z M^2_Z\right]} (e_\ell e_q) 
%\\  \nonumber
% &\times  
 \left[g_V^{Z \ell} 
 g_{V }^{Zq} 
 (1 + \cos^2\theta^*) + 2 g_A^{Z \ell} g_{A }^{Z q} \cos\theta^*\right]
 \\ \nonumber
 &+ \frac{M^4_{\ell\ell}}{16 \sin^4\theta_W \cos^4\theta_W\left[(M^2_{\ell\ell} - M^2_Z)^2 + \Gamma^2_Z M^2_Z\right]} 
\\  \nonumber
 &\times 
 \left\{  [ ( g_A^{Z \ell } )^2 + ( g_V^{Z \ell })^2  ]  
 [ ( g_{A }^{Z q } )^2 + (g_{V }^{Z q } )^2 ] 
 (1+\cos^2\theta^*) 
%  \right. 
%\\   \nonumber 
%& 
%\left.  
 + 8 
 g_A^{Z \ell }  g_V^{Z \ell }  
 g_{A }^{Z q} g_{V }^{Z q} \cos\theta^* \right\} . \nonumber
\end{align}
Here $M_Z$ and $\Gamma_Z$ are the mass and the width of the $Z$ boson, $e_\ell$ and $e_q$ are the lepton and quark electric charges, 
$g_V^{Z \ell }=-1/2 + 2 \sin^2 \theta_W$ and $g_A^{Z \ell }=-1/2$ are the vector and axial couplings of leptons, 
and 
$g_V^{Z q }$ and $g_A^{Z q }$ are the SMEFT 
vector and axial couplings of quarks in Eqs.~(\ref{gLR_SMEFT}), (\ref{gVA_SMEFT}). 
The first and third terms on the right hand side of 
Eq.~(\ref{eq:propagators}) are the square of the $s$-channel diagrams with 
photon and $Z$-boson mediators respectively, while the second term is 
the interference between the two.

The forward-backward asymmetry $A_{\rm{FB}}^*$ is defined as 
\begin{equation}
  A_{\rm{FB}}^* = \frac { d^2 \sigma / d M_{\ell \ell} d y_{\ell\ell} [\cos\theta^*>0] - 
  d^2 \sigma / d M_{\ell \ell} d y_{\ell\ell} [\cos\theta^*<0] }    
    { d^2 \sigma / d M_{\ell \ell} d y_{\ell\ell}  [\cos\theta^*>0] 
    + d^2 \sigma / d M_{\ell \ell} d y_{\ell\ell} [\cos\theta^*<0] }.
  \label{eq:afb}
\end{equation}
We will consider the measurement of the $A_{\rm{FB}}^*$ asymmetry 
differentially in $M_{\ell\ell}$ and $y_{\ell\ell}$  
according to Eqs.~(\ref{eq:triple_diff}), (\ref{eq:afb}).

To perform this study, 
we extend the implementation~\cite{Accomando:2019vqt} of 
the $A_{\rm{FB}}^*$ asymmetry 
in the  \texttt{xFitter} platform~\cite{Alekhin:2014irh,xFitter:2022zjb} 
to i) include the SMEFT couplings described above 
in Eqs.~(\ref{gLR_SMEFT}), (\ref{gVA_SMEFT}),
and ii) upgrade the calculations to double-differential 
distributions in both invariant mass $M_{\ell\ell}$ and rapidity $y_{\ell\ell}$ of the 
di-lepton final-state system. 
The collider energy, acceptance cuts and bin
 boundaries in $M_{\ell\ell}$ and  $y_{\ell\ell}$  are adjustable 
parameters in the present computation. 
%We take acceptance cuts for the lepton 
%pseudorapidity and transverse momentum 
%$ | \eta_\ell | <  5 $ and  $p_T^\ell > 20 $ GeV. 
Fiducial selections are applied to the leptons, by requiring them to have a transverse momentum $p_T^\ell > 20 $ GeV and pseudorapidity $ | \eta_\ell | <  5 $.
The mass effects of charm and bottom quarks in the matrix element
 are neglected, as appropriate for a high-scale process, and the 
calculation is performed in the  $n_f = 5$ flavour scheme.  
The input theoretical parameters are chosen to be 
 the ones from the EW $G_\mu$ scheme, which minimizes the impact of NLO EW corrections, see e.g.\ Ref.~\cite{Dittmaier:2009cr}.
 The explicit values for the relevant parameters in our 
analysis are the following: $M_Z = 91.188$ GeV,  $\Gamma_Z = 2.441 $ GeV,
$M_W = 80.149 $ GeV, $\alpha_{em} = 1/132.507$. 

\begin{figure}[!htbp]
    \centering
    \includegraphics[width=1.00\textwidth]{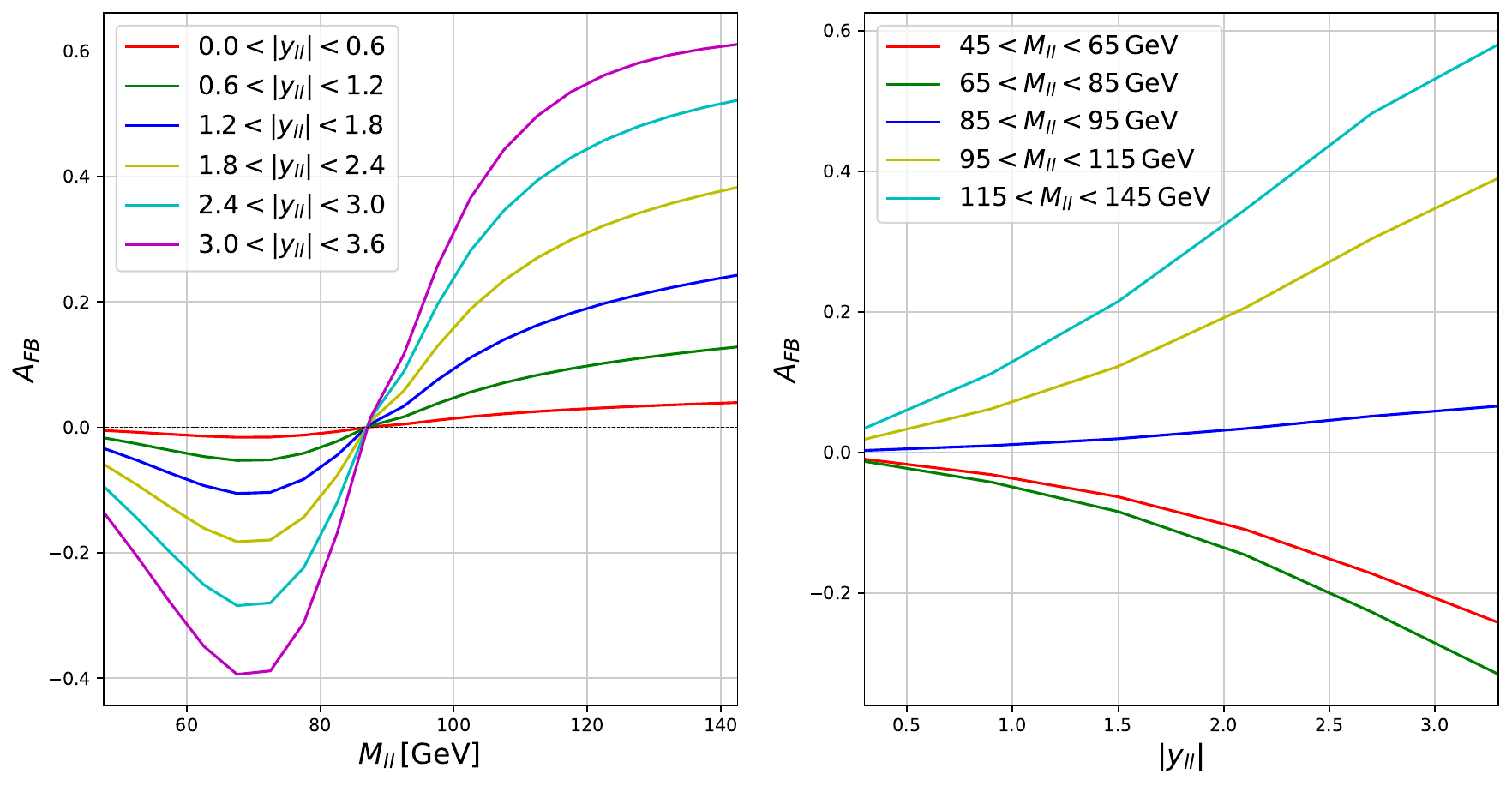}
    \caption{The predicted \afb as a function of the invariant mass of the dilepton system in different rapidity intervals (left) and rapidity of the dilepton system in different invariant mass intervals (right) at LO in the SM.}
    \label{fig:afb}
\end{figure}

The predicted $A_{\rm{FB}}$   as a function of the invariant
 mass and rapidity of the di-lepton system at LO in the SM is shown in 
Fig.~\ref{fig:afb}. 
The   $A_{\rm{FB}}$  
 crosses zero  around   $  M_{\ell\ell} \approx M_Z $. Also, 
due to its definition using the longitudinal boost
 of the di-lepton system, it approaches zero at $y_{\ell\ell} = 0$.
For this calculation, we used the HERAPDF2.0~\cite{H1:2015ubc} PDF set, however, its general features do not depend on the PDF set.%
\footnote{Theoretical predictions for $A_{\rm{FB}}$ obtained using other PDF sets can be found e.g.\ in Ref.~\cite{ATLAS:2018gqq} or Ref.~\cite{Accomando:2017scx}.} 
\neww{We do not include any QED effects in our calculation since the experimental data are typically corrected for QED effects, and the uncertainties in these corrections are much smaller than the PDF uncertainties in \afb~\cite{ATLAS:2018gqq}.}

\begin{figure}[!htbp]
    \centering
    \includegraphics[width=1.00\textwidth]{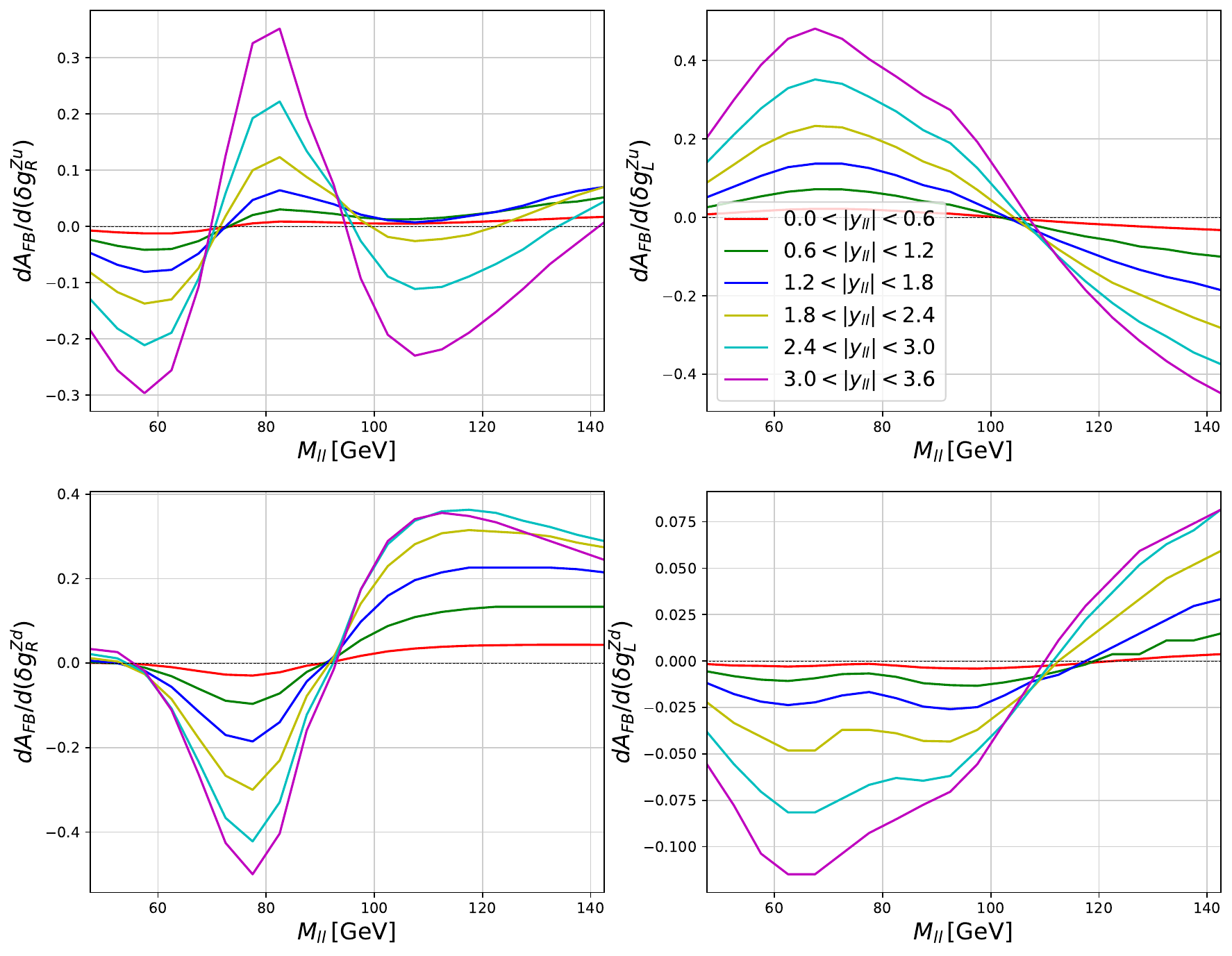}
    \caption{The partial derivatives of the predicted \afb with respect to $\delta g_R^{Zu}$ (upper left), $\delta g_L^{Zu}$ (upper right), $\delta g_R^{Zd}$ (lower left) and $\delta g_L^{Zd}$ (lower right) couplings as a function of the invariant mass of the dilepton system in different rapidity intervals at LO.}
    \label{fig:afb-deriv-m}
\end{figure}

\begin{figure}[!htbp]
    \centering
    \includegraphics[width=1.00\textwidth]{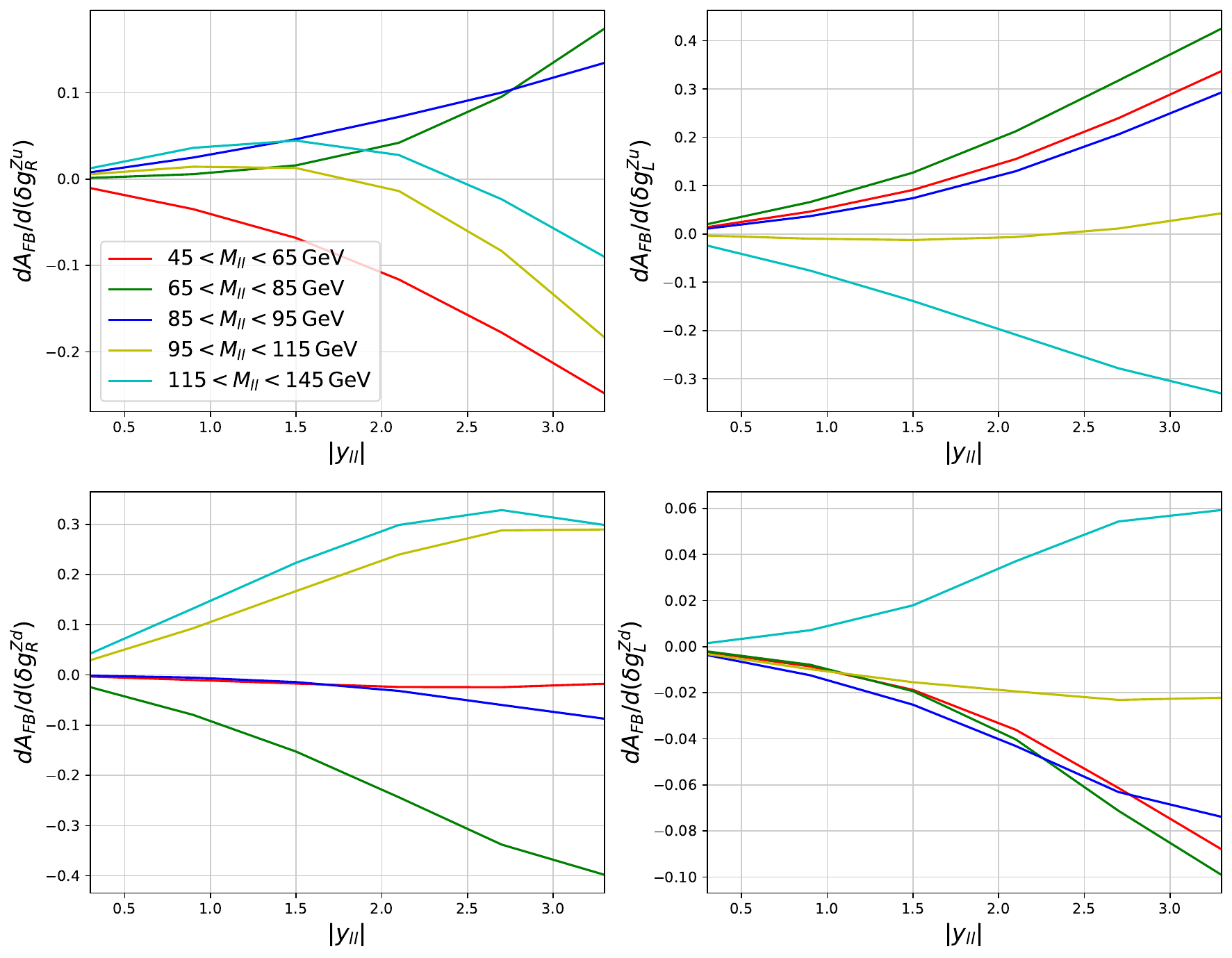}
    \caption{The partial derivatives of the predicted \afb with respect to $\delta g_R^{Zu}$ (upper left), $\delta g_L^{Zu}$ (upper right), $\delta g_R^{Zd}$ (lower left) and $\delta g_L^{Zd}$ (lower right) couplings as a function of the rapidity of the dilepton system in different invariant mass intervals at LO.}
    \label{fig:afb-deriv-y}
\end{figure}

\begin{figure}[!htbp]
    \centering
    \includegraphics[width=1.00\textwidth]{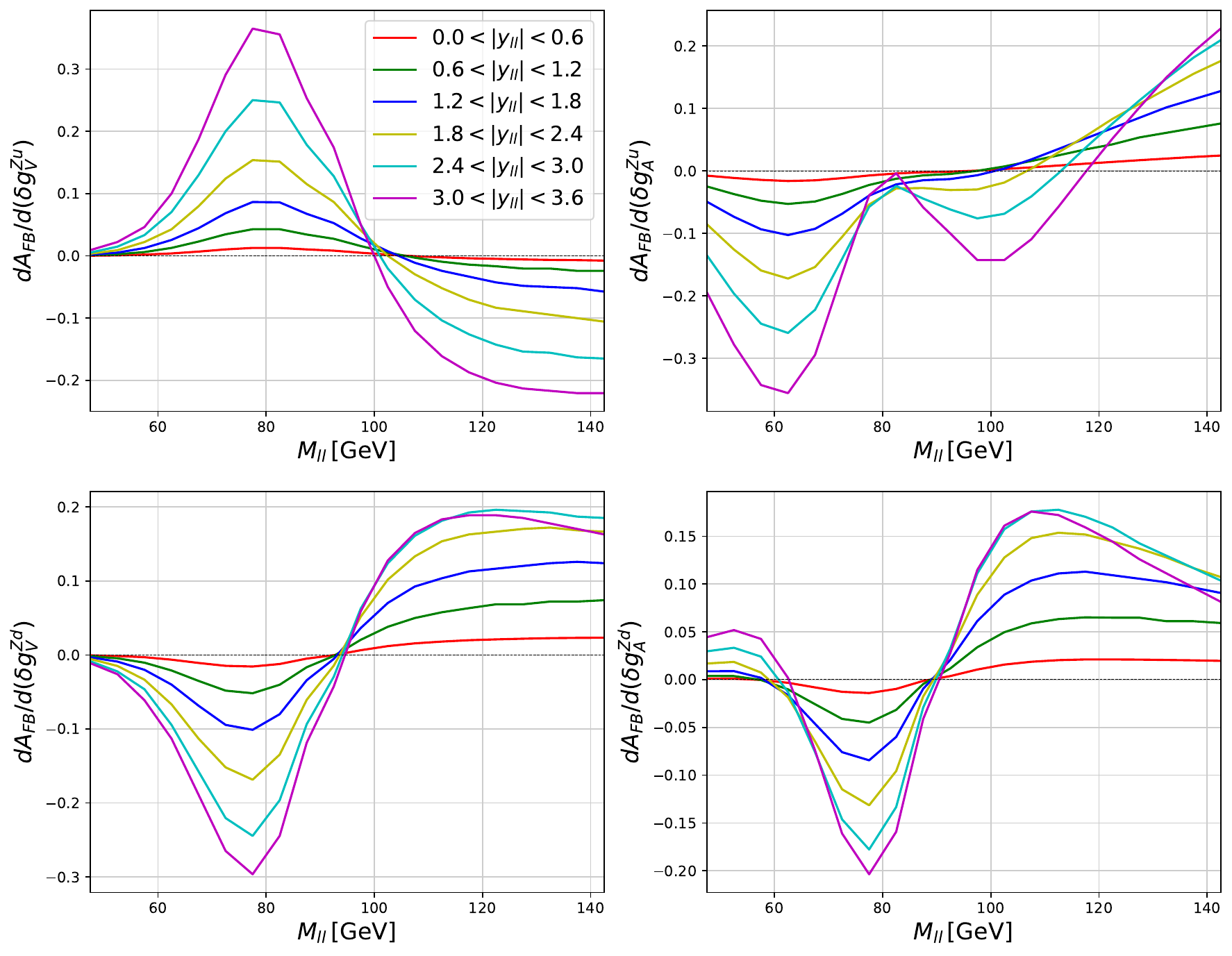}
    \caption{Same as in Fig.~\ref{fig:afb-deriv-m} for the axial and vector couplings.}
    \label{fig:afb-deriv-m-av}
\end{figure}

\begin{figure}[!htbp]
    \centering
    \includegraphics[width=1.00\textwidth]{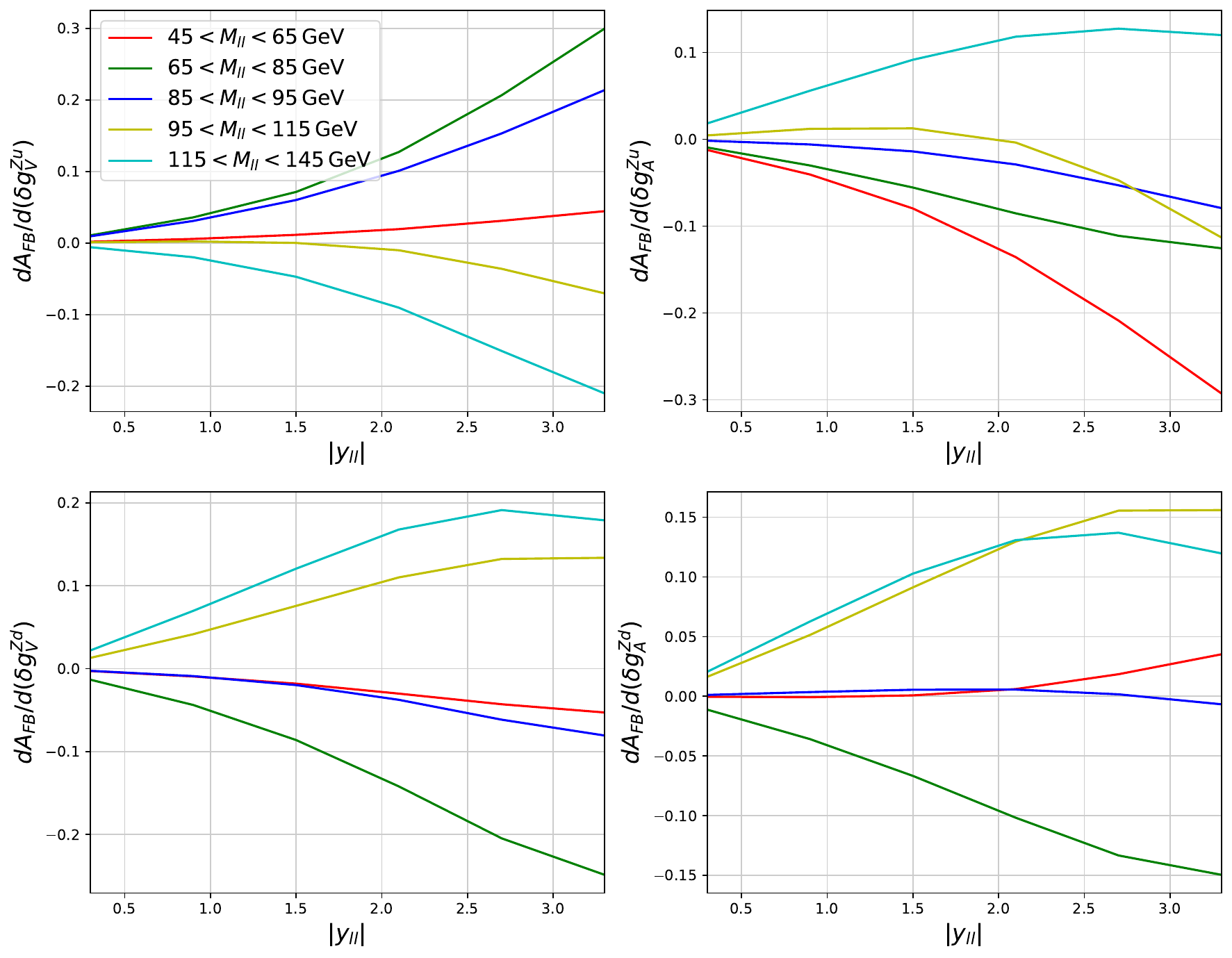}
    \caption{Same as in Fig.~\ref{fig:afb-deriv-y} for the axial and vector couplings.}
    \label{fig:afb-deriv-y-av}
\end{figure}

We next investigate the dependence of the predicted $A_{\rm{FB}}$ 
on the couplings. In Figs.~\ref{fig:afb-deriv-m} and  \ref{fig:afb-deriv-y} 
we show the numerically-calculated partial derivatives of the 
$A_{\rm{FB}}$ with respect to each 
coupling as a function of the invariant mass and rapidity of the dilepton
 system.  
Furthermore, in Figs.~\ref{fig:afb-deriv-m-av} and  \ref{fig:afb-deriv-y-av} these derivatives are shown with respect to the axial and vector couplings.
It is instructive to see from Fig.~\ref{fig:afb-deriv-m} that the partial derivatives as  functions of $ M_{\ell\ell}$  cross zero
 at  values of $ M_{\ell\ell}$  which are almost independent of $y_{\ell\ell}$. As a result,
 the partial derivatives as  functions of  $y_{\ell\ell}$  vanish 
after integrating over the  $ M_{\ell\ell}$     
 regions which contain such turnover points near their centers (e.g.,  
$ \partial  A_{\rm{FB}} / \partial  \delta  g_{R }^{Zd}  \approx 0$  for
 85 $ <   M_{\ell\ell}  < 95 $ GeV). 
This is an important observation for experimental analyses which
 aim to measure $A_{\rm{FB}}$ in bins of $ M_{\ell\ell}$  and $y_{\ell\ell}$: in 
particular, in order to retain sensitivity   
 to the couplings, the binning scheme should be chosen carefully, preferably such that the
 points where the derivatives vanish are placed at the bin boundaries, rather than at their
 centers. Furthermore, the magnitudes of the derivatives give an idea of which phase-space
 regions are expected to be most sensitive to the couplings. However, one needs to
 take into account the expected statistical uncertainties also. Therefore, we 
will come back to this after introducing the pseudodata in the next section. 

\section{Generation of pseudodata sets}
\label{sec:pseudodata} 

Suitable data files which mimic future measurements at the HL-LHC have been generated for the analysis.
\new{Namely, we used the expected HL-LHC luminosity, SM theoretical predictions and our assumption of 20\% for the detector response to predict the number of events and statistical uncertainties for the future AFB measurement at the HL-LHC.}
The central values of the pseudodata points are set to the SM theoretical predictions.
An important piece of information contained in the data files is the statistical precision associated to the \afb experimental measurements in each bin.
It is given by:
\begin{equation}
  \Delta \afb = \sqrt{\frac{1-{\afb}^2}{N}}, 
  \label{eq:afb_error}
\end{equation}
\noindent
where $N$ is the expected total number of events in a specific invariant mass interval.
We use the number of events with electron pairs from $Z$ decays as predicted at LO with the acceptance cuts $ | \eta_\ell | <  5 $ and  $p_T^\ell > 20 $ GeV and introduce a further correction factor of $20\%$ to model a realistic detector response~\cite{CMS:2014lcz}.
The choice of LO accuracy for the expected number of events provides a conservative estimation of the statistical uncertainty. 
The higher-order QCD corrections through next-to-next-to-leading order (NNLO) for the DY process are, generally, moderate and do not distort much differential distributions, see, e.g. Ref.~\cite{Alekhin:2021xcu}. 
We have checked that the usage of NNLO QCD predictions would increase the expected number of events by factor $1.1\text{--}1.4$ depending on the phase space region, so the statistical uncertainties does not change by more than 20\%~\cite{Accomando:2019vqt}. 
Furthermore, we have tested our approach by comparing the statistical uncertainties from the ATLAS measurement of \afb~\cite{ATLAS:2018gqq} with the ones produced using our pseudodata scenario, and found a reasonable agreement within a factor of two.\footnote{Since the results of the ATLAS measurement~\cite{ATLAS:2018gqq} are reported in the full phase space of the leptons, the statistical uncertainties depend also on the extrapolation factors which we did not include in this study.}

The pseudodata have been generated for the collider centre-of-mass energy of 13 TeV and integrated luminosity of 3000 fb$^{-1}$, the designed integrated luminosity at the end of the HL-LHC stage~\cite{Gianotti:2002xx}.
To explore different proton PDF sets, several data files have been generated adopting the recent NNLO variants of the PDF sets CT18~\cite{Hou:2019efy}, NNPDF4.0~\cite{NNPDF:2021njg}, ABMP16~\cite{Alekhin:2017kpj}, HERAPDF2.0~\cite{H1:2015ubc} and MSHT20~\cite{Bailey:2020ooq} along with their respective uncertainties as provided by each fitting group.

Theoretical uncertainties arising from the choice of factorization and renormalization scales have been assessed.
For this purpose, we used theoretical predictions at NLO obtained using \texttt{MadGraph5\_aMC@NLO}~\cite{Alwall:2014hca} interfaced to \texttt{APPLgrid}~\cite{Carli:2010rw} through \texttt{aMCfast}~\cite{Bertone:2014zva}. We have found that the impact of scale variations by the conventional factor of two in the theoretical predictions at NLO is small compared to the statistical uncertainties of the pseudodata, and thus we do not include it in our analysis. A similar study (focused on the impact of the \afb on PDFs) was performed in Ref.~\cite{Accomando:2019vqt}. Also, it is worth mentioning that even smaller theoretical uncertainties could be expected at NNLO in QCD.
Furthermore, while it would be important to include the NLO EW effects in an analysis of experimental data aiming to obtain accurate central values, they are not expected to bring significant modification to the uncertainties.

\begin{figure}[!htbp]
    \centering
    \includegraphics[width=0.49\textwidth]{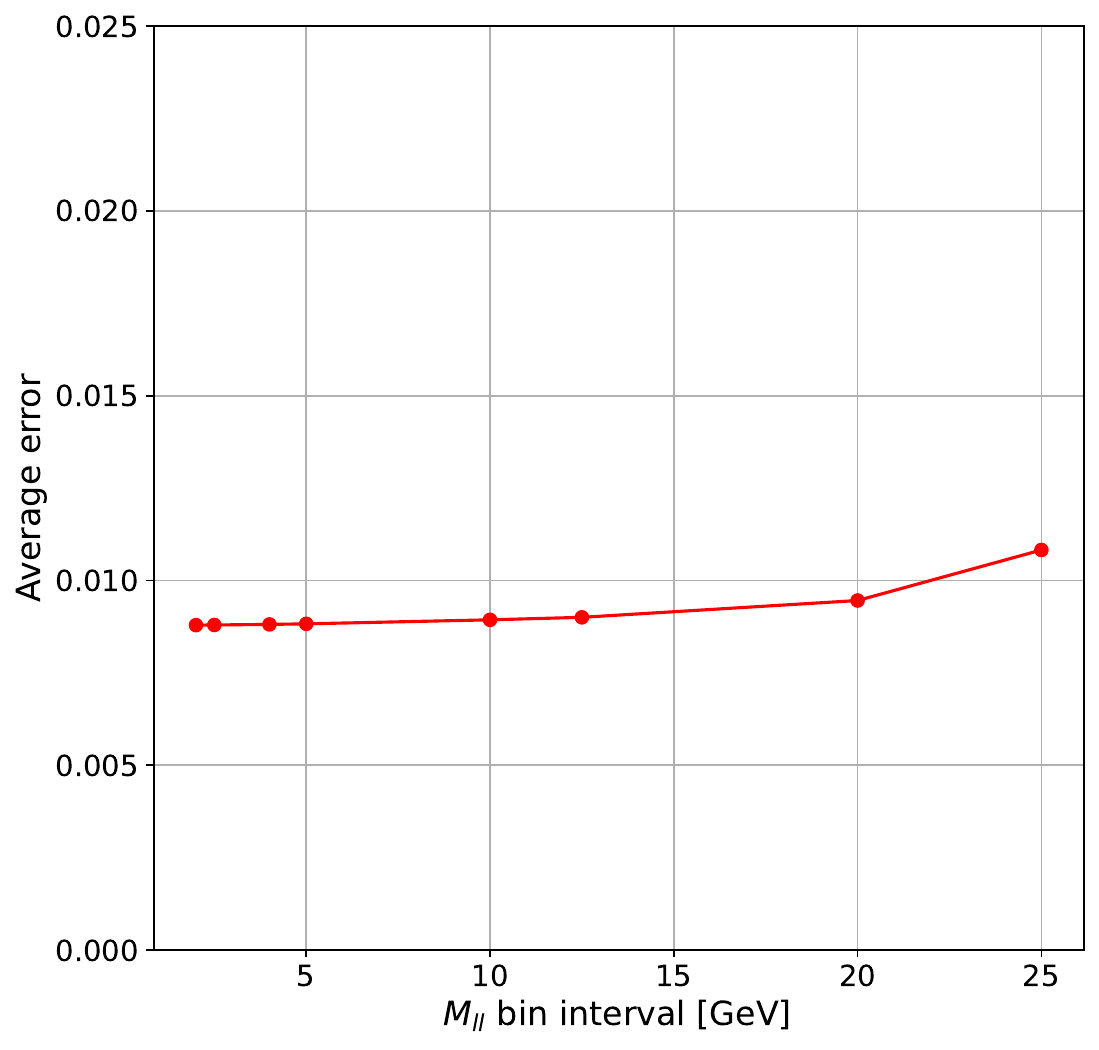}
    \includegraphics[width=0.49\textwidth]{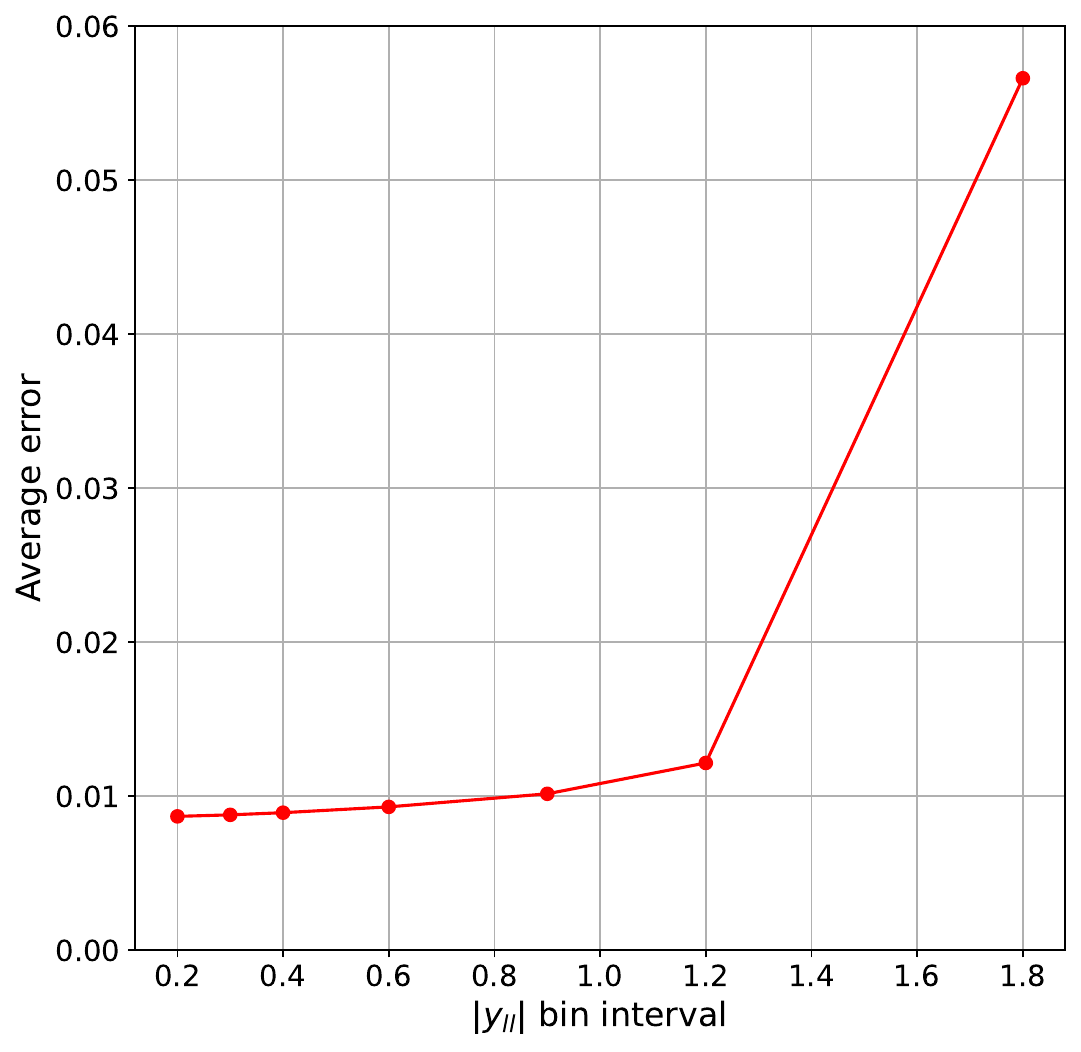}
    \caption{The average error in the fitted couplings as a function of the invariant mass (left) and rapidity (right) bin widths.}
    \label{fig:vol}
\end{figure}

Another important ingredient of the pseudodata is the binning scheme. As discussed in Section~\ref{sec:xfitter}, bins with the turnover points of the partial derivatives of \afb with respect to the couplings should be avoided, because in such bins the sensitivity to the couplings is washed out after integrating over the bin. 
In general, one needs as fine as possible bins in order to maximize the sensitivity to the parameters of interest. 
However, due to limited detector resolution too fine bins cannot be used. We have optimized the bin widths based on the precision of the fitted couplings which we extract from the pseudodata. 
As a figure of merit, we have used the geometrical average error of the couplings, i.e.\ the fourth root of the product of the resulting uncertainties on each of the four determined couplings, $\sqrt[4]{\Delta\delta g_R^{Zu}\Delta\delta g_R^{Zd}\Delta\delta g_L^{Zu}\Delta\delta g_L^{Zd}}$. 
In Fig.~\ref{fig:vol} we show this quantity as a function of the invariant mass and rapidity bin widths. 
Based on this study, we chose the bin width of $5$~GeV in the invariant mass and $0.6$ in the rapidity of the dilepton system, since 
a further reduction of the bin width does not improve the sensitivity to the couplings significantly. 
\new{Namely, using the 10 GeV bin width in $M_{ll}$, one will get about 1\% larger uncertainties on the fitted couplings than using the 5 GeV bin width, which we find already sizeable. On the other hand, using a smaller bin width $<5$ GeV provides only a marginal further improvement at permil level.
These bin widths are feasible given the resolution of the existing detectors~\cite{ATLAS:2022jjr,ATLAS:2018krz}.}
The minimum expected number of events in a bin amounts to $\sim 10000$, thus the probability distribution function of the statistical uncertainty is well approximated by a normal distribution.
\new{Our binning scheme is given in Table~\ref{tab:bins}.}

\begin{table}[htb]
\centering
\begin{tabular}{lll}
    Observable& Bin boundaries \\
    \hline
    $M_{ll}$ [GeV] & 45,50,55,60,65,70,75,80,85,90,95,100,105,110,115,120,125,130,135,140,145\\
    $|y_{ll}|$ & $0,0.6,1.2,1.8,2.4,3.0,3.6$\\
\end{tabular}
\caption{The binning scheme used in our analysis.}
\label{tab:bins}
\end{table}

\begin{figure}[!htbp]
    \centering
    \includegraphics[width=1.00\textwidth]{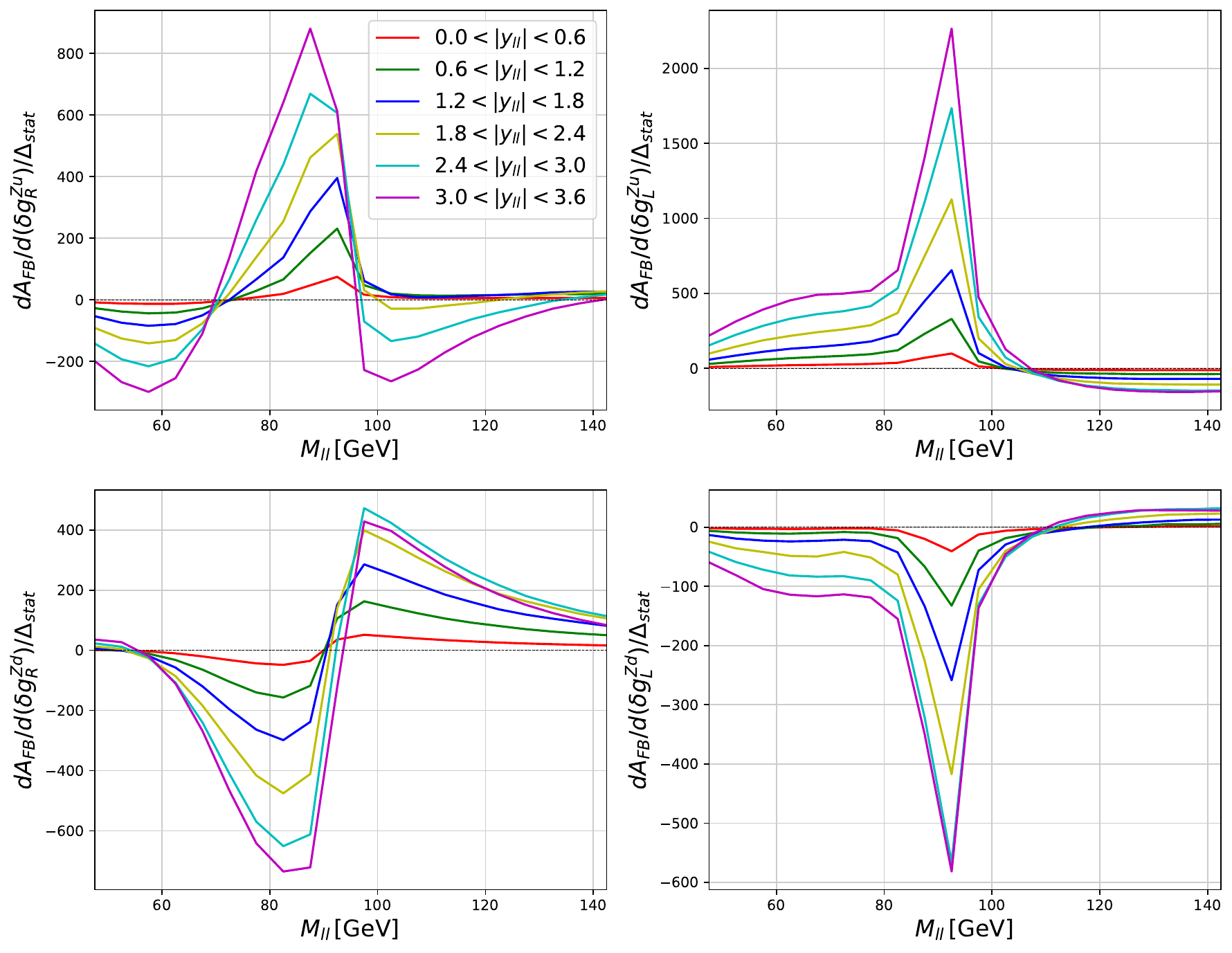}
    \caption{The partial derivatives of the predicted \afb with respect to $\delta g_R^{Zu}$ (upper left), $\delta g_L^{Zu}$ (upper right), $\delta g_R^{Zd}$ (lower left) and $\delta g_L^{Zd}$ (lower right) couplings weighted by the inverse of the statistical uncertainty as a function of the invariant mass of the dilepton system in different rapidity intervals at LO.}
    \label{fig:afb-deriv-unc-m}
\end{figure}

\begin{figure}[!htbp]
    \centering
    \includegraphics[width=1.00\textwidth]{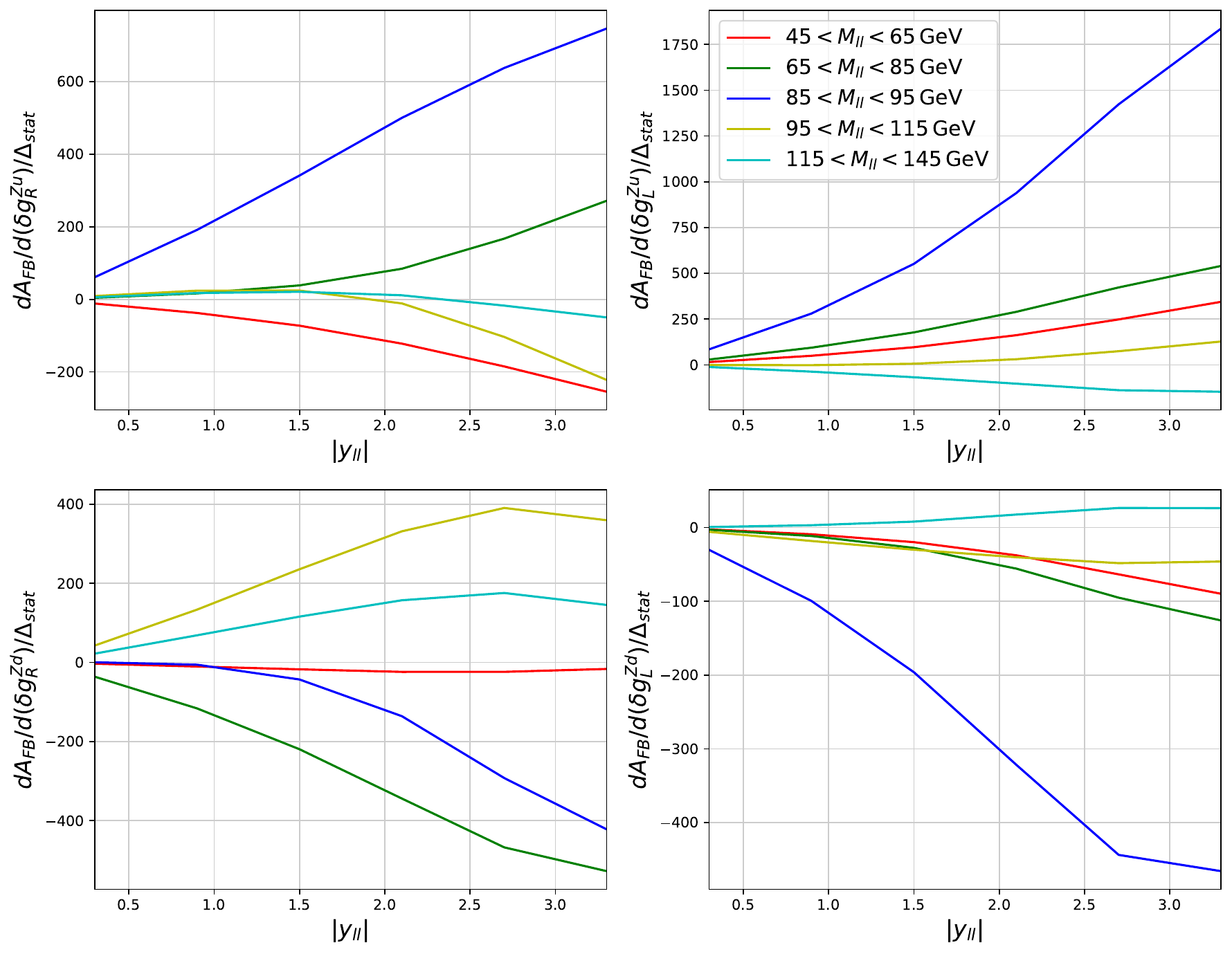}
    \caption{The partial derivatives of the predicted \afb with respect to $\delta g_R^{Zu}$ (upper left), $\delta g_L^{Zu}$ (upper right), $\delta g_R^{Zd}$ (lower left) and $\delta g_L^{Zd}$ (lower right) couplings weighted by the inverse of the statistical uncertainty as a function of the rapidity of the dilepton system in different invariant mass intervals at LO.}
    \label{fig:afb-deriv-unc-y}
\end{figure}

\begin{figure}[!htbp]
    \centering
    \includegraphics[width=1.00\textwidth]{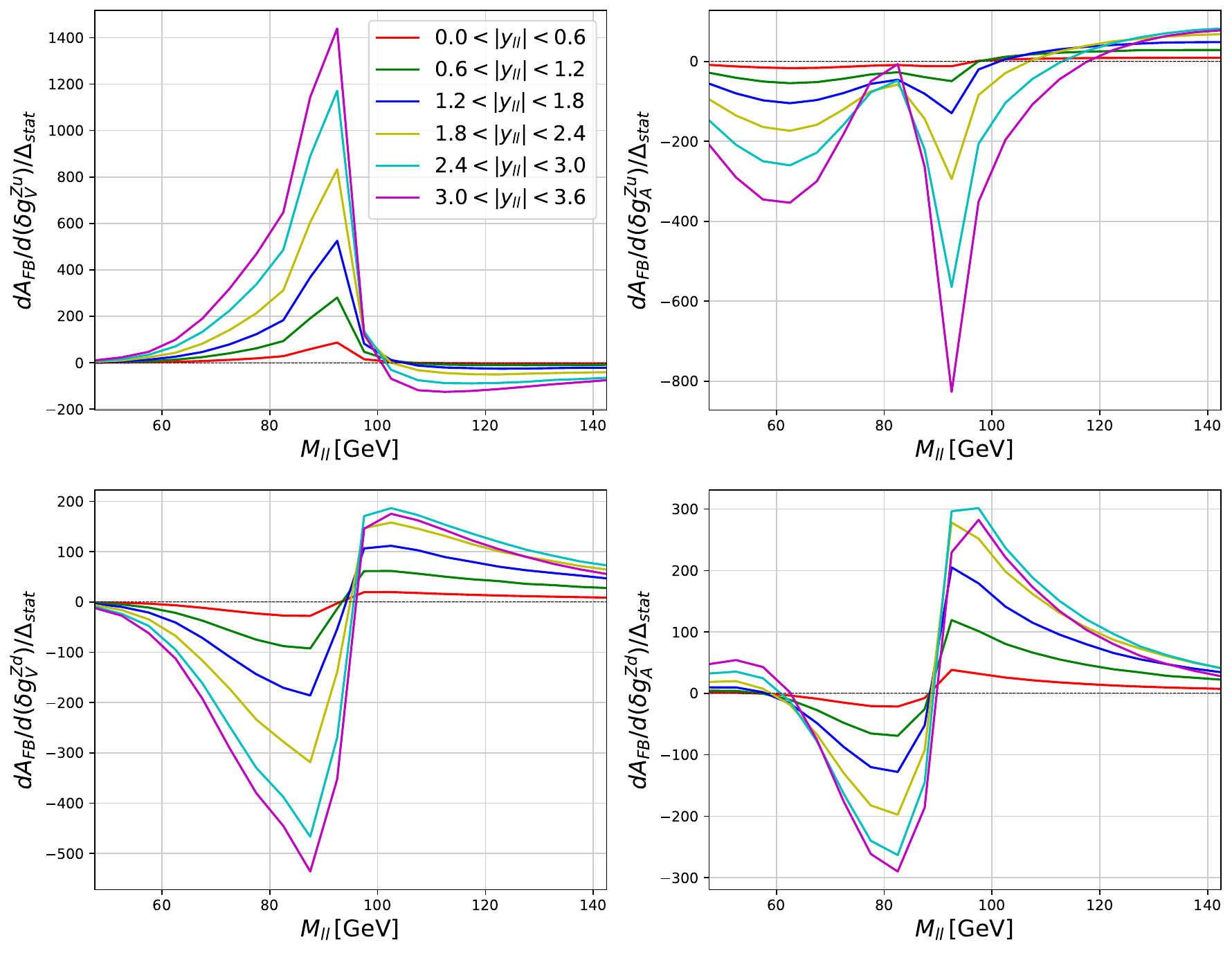}
    \caption{Same as in Fig.~\ref{fig:afb-deriv-unc-m} for the axial and vector couplings.}
    \label{fig:afb-deriv-unc-m-av}
\end{figure}

\begin{figure}[!htbp]
    \centering
    \includegraphics[width=1.00\textwidth]{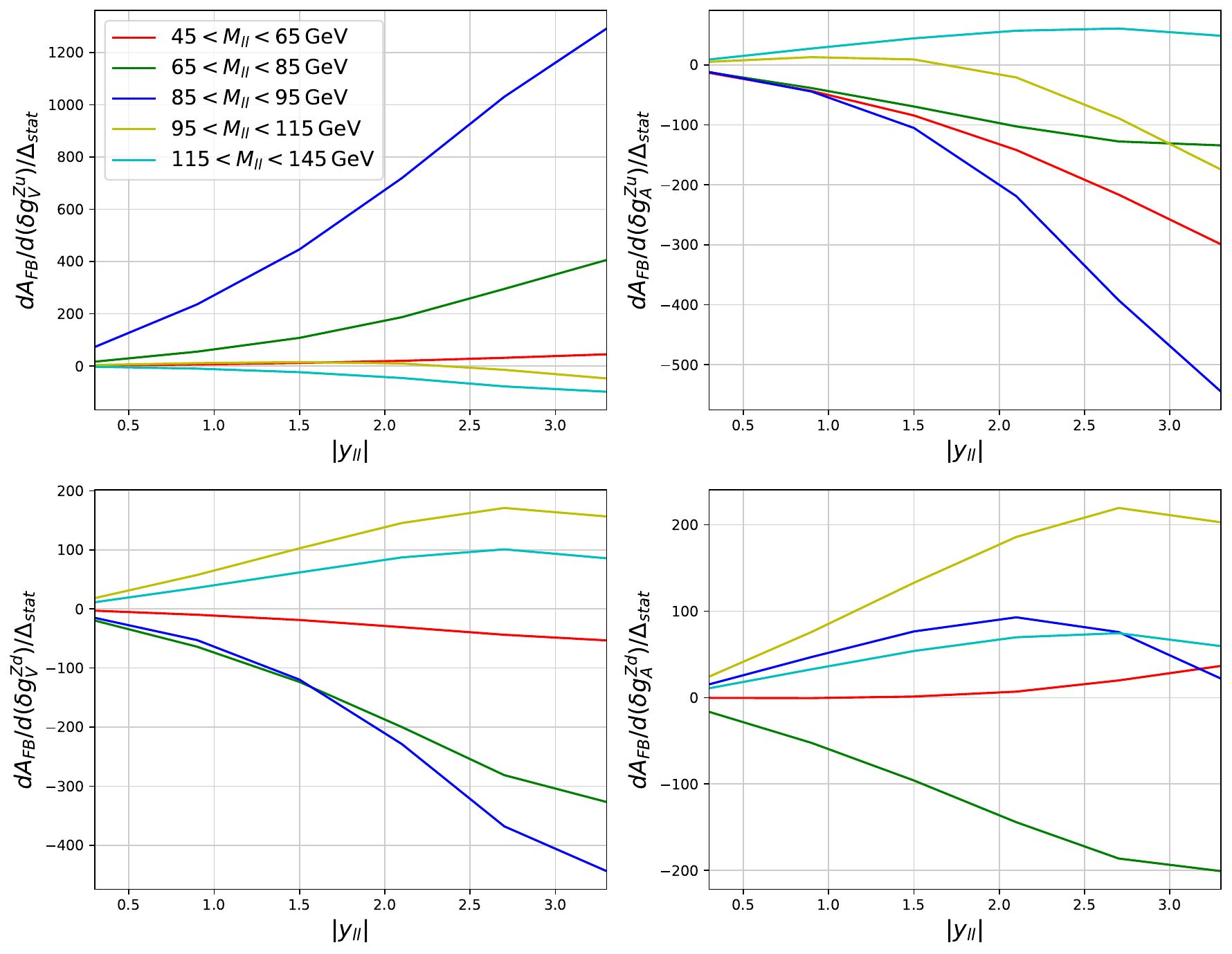}
    \caption{Same as in Fig.~\ref{fig:afb-deriv-unc-y} for the axial and vector couplings.}
    \label{fig:afb-deriv-unc-y-av}
\end{figure}

It is illustrative to look at the magnitudes of the partial derivatives as a function of the invariant mass and rapidity of the dilepton system weighted by the inverse of the statistical uncertainty of pseudodata in the corresponding phase-space region, as shown in Figs.~\ref{fig:afb-deriv-unc-m}--\ref{fig:afb-deriv-unc-y-av}. This quantity is proportional to the sensitivity to the couplings which can be extracted from such a phase-space region. The largest sensitivity to the couplings is expected in the region $55 \lesssim M_{ll} \lesssim 110$~GeV (see Figs.~\ref{fig:afb-deriv-unc-m}, \ref{fig:afb-deriv-unc-m-av}), and in our analysis we have adopted a slightly wider range $45 < M_{ll} < 145$~GeV. 
\neww{Possible contributions from four-fermion operators in this kinematic range are expected at the $\lesssim 1\permil$ level~\cite{Boughezal:2023nhe}. As a simplifying assumption, we neglected them in our analysis. A study of their possible impact is described in Appendix~\ref{sec:4f}.}

We have limited the rapidity region $|y_{ll}|<3.6$ assuming the extension of the detector acceptance up to pseudorapidity $|\eta_l|<5$ (while potentially the region $|y_{ll}|>3.6$ could provide even further improvement for the sensitivity to the couplings).

\section{Results}
\label{sec:results}

The pseudodata sets are fitted with the four modifications to the couplings $\delta g_L^{Zu}$, $\delta g_R^{Zu}$, $\delta g_L^{Zd}$, $\delta g_R^{Zd}$ being free parameters. \new{The fit is performed using the \texttt{MINUIT}~\cite{James:1975dr} library by minimizing a \chisq expression:
\begin{equation} 
	\label{eq:chisq}
    \chisq = \Sigma_i\frac{(m^i-\Sigma_j\gamma^i_jm^is_j-\mu_i)^2}{\delta_i^2}+\Sigma_js_j^2,
\end{equation} 
which follows the one used in Ref.~\cite{H1:2015ubc}.
Here, $\mu^i$ is the measured value in bin $i$, $\delta_i$ is its experimental uncertainty, $m^i$ is the theoretical prediction, $\gamma^i_j$ is its relative uncertainty due to the PDF eigenvector $j$ which is shifted in units of sigma by $s_j$.
This treatment of the PDF uncertainties follows the so-called profiling technique~\cite{Paukkunen:2014zia,HERAFitterdevelopersTeam:2015cre}. \newepjc{In this method, the PDF uncertainties are included in the \chisq using nuisance parameters $s_j$ which are further constrained according to the tolerance criterion of the fit. The number of nuisance parameters $s_j$ corresponds to the number of eigenvectors for each PDF set. While $s_j$ are free parameters when minimizing the \chisq, they do not change the number of degrees of freedom, because for each $s_j$ the corresponding prior is added, represented by the last term in Eq.~(\ref{eq:chisq}), which acts as a data point.}} As the tolerance criterion we use $\Delta\chisq=1$. In this approach, one assumes that the new data are compatible with the theoretical predictions using the existing PDF set. No further theoretical uncertainties beyond the PDF uncertainties are considered when calculating the \chisq. The uncertainties on the fitted parameters are obtained using the \texttt{HESSE} method which computes numerically the second derivatives of the \chisq with respect to the fitted parameters~\cite{Alekhin:2014irh,xfittergitlab}. This assumes that the dependence of the theoretical predictions on the parameters of interest is linear near the minimum of the \chisq. We cross-checked the uncertainties using the \texttt{MINOS} algorithm which uses the profile likelihood method to compute asymmetric confidence intervals, as well as the \texttt{MNCONT} algorithm~\cite{James:1975dr} which explicitly finds 2D contours where the \chisq is minimal, and found a good agreement with the hessian uncertainties.

\begin{figure}[!htbp]
    \centering
    \includegraphics[width=1.00\textwidth]{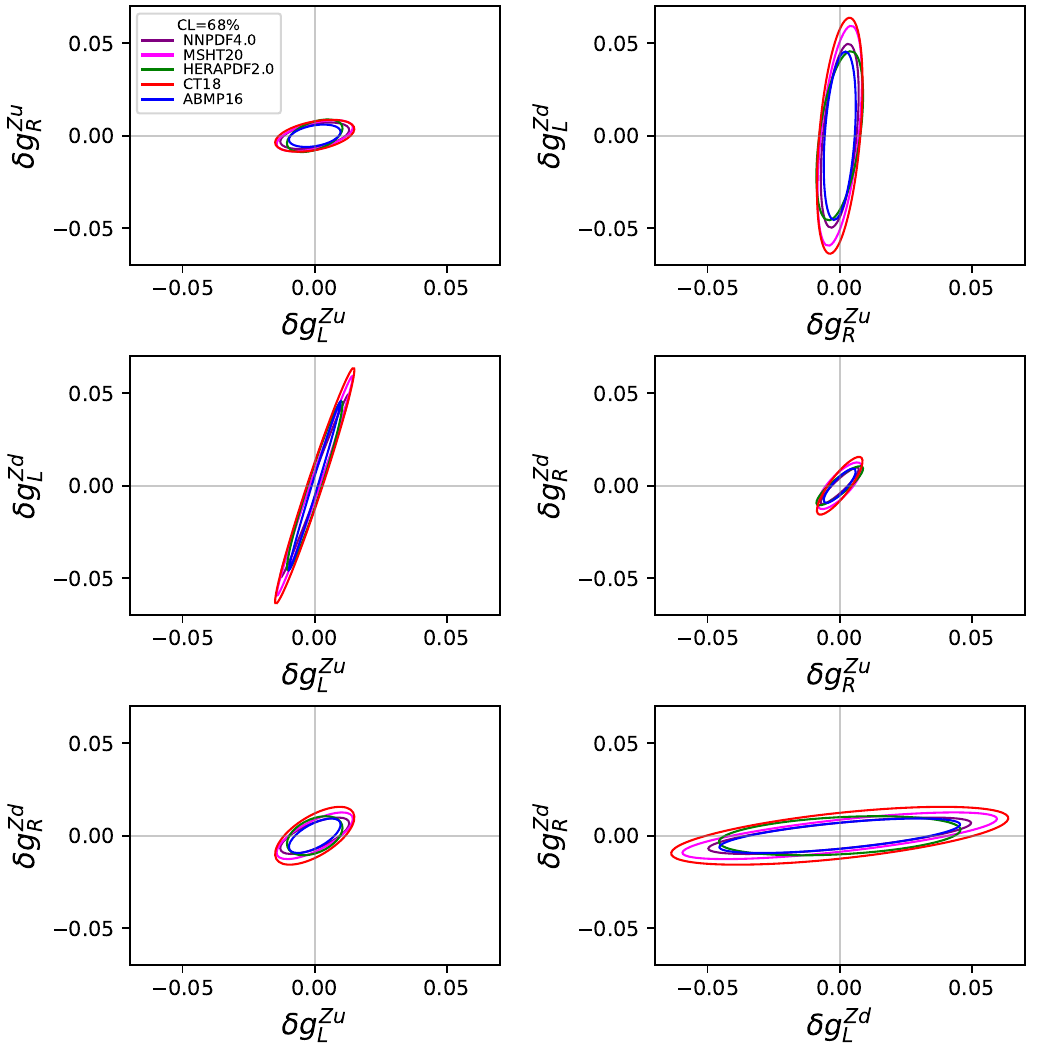}
    \caption{Allowed regions for all pairs of corrections to the $Z$ couplings to $u$- and $d$-type quarks obtained using different PDF sets.}
    \label{fig:ell_pdf}
\end{figure}

\begin{figure}[!htbp]
    \centering
    \includegraphics[width=1.00\textwidth]{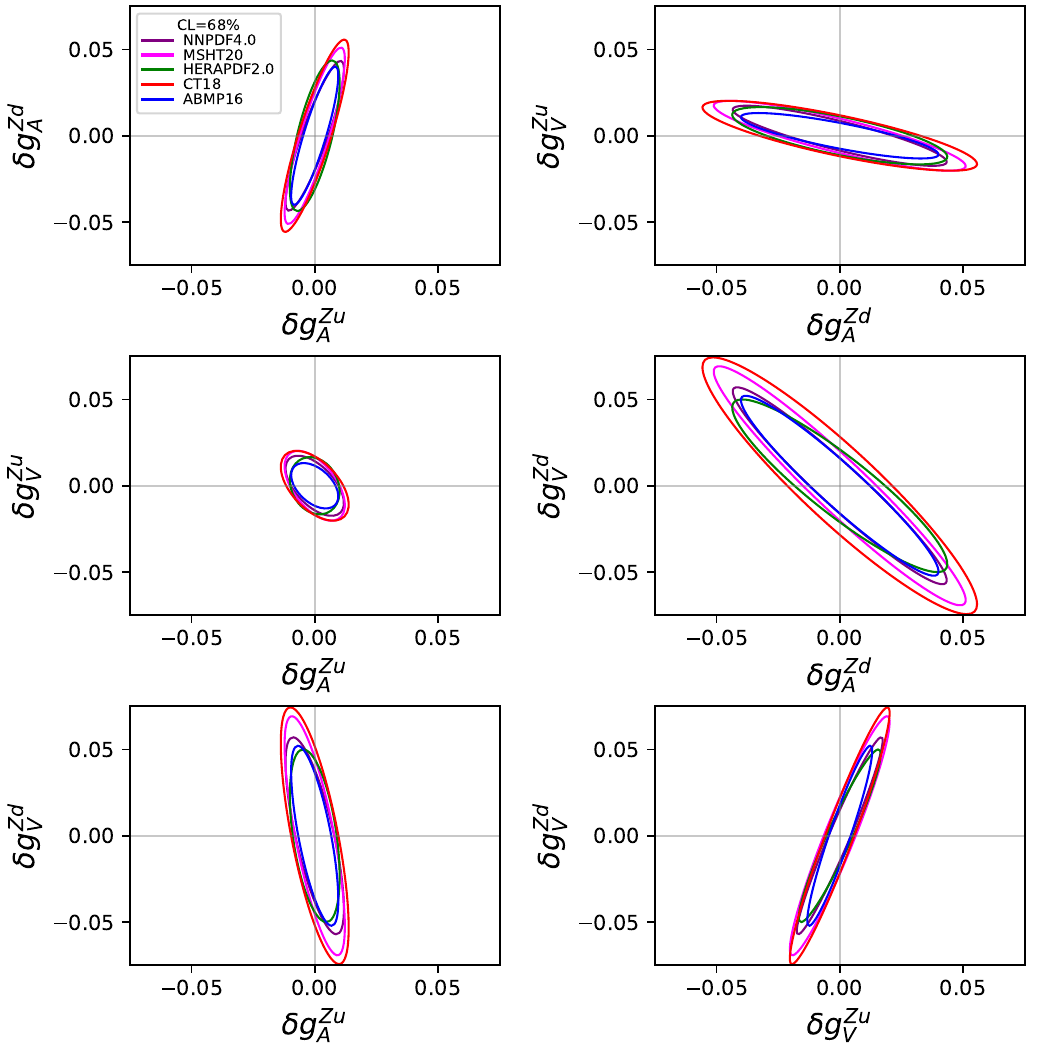}
    \caption{Same as in Fig.~\ref{fig:ell_pdf} for the axial and vector couplings.}
    \label{fig:ell_pdf_av}
\end{figure}

\begin{figure}[!htbp]
    \centering
    \includegraphics[width=1.00\textwidth]{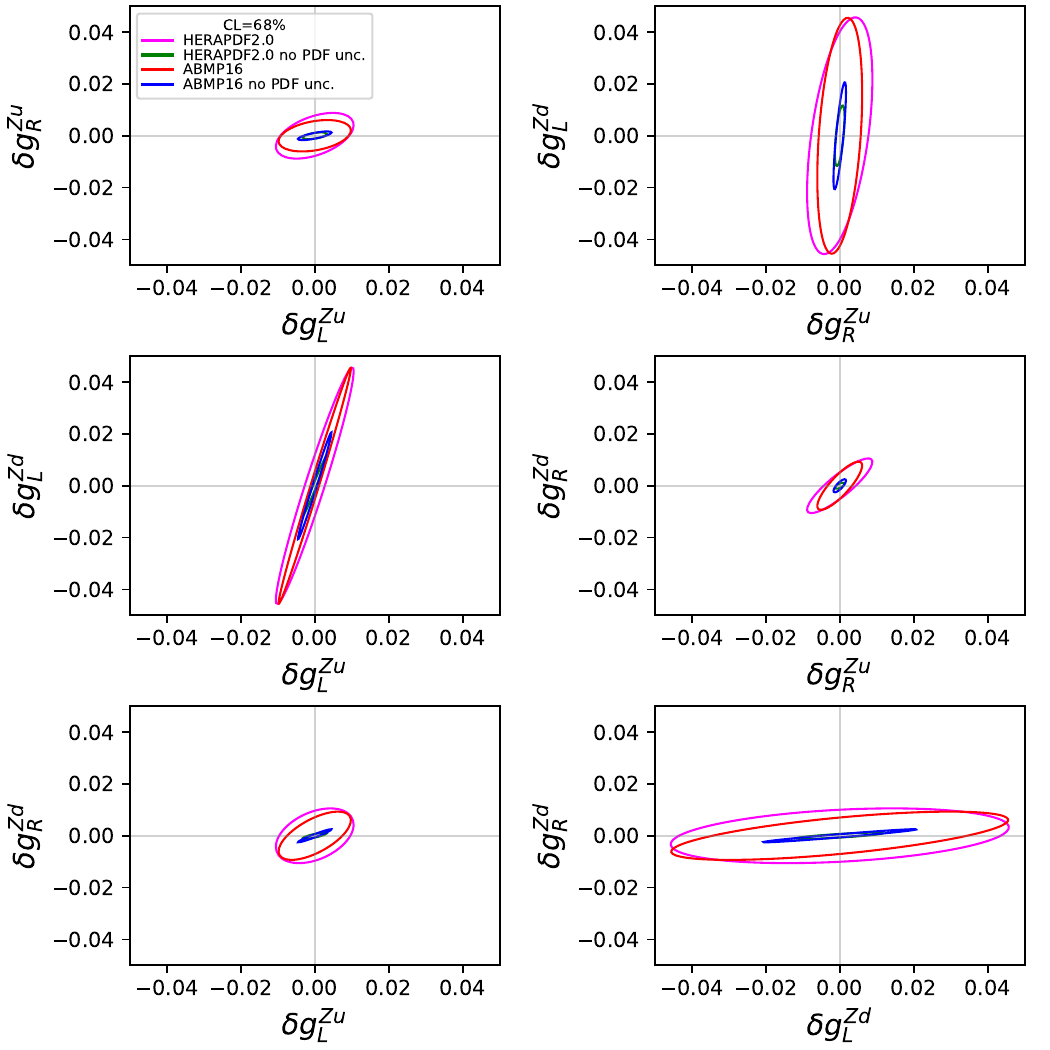}
    \caption{Allowed regions for all pairs of corrections to the $Z$ couplings to $u$- and $d$-type quarks obtained using the ABMP16 and HERAPDF2.0 PDF sets with and without PDF uncertainties.}
    \label{fig:ell_pdf_unc}
\end{figure}

\begin{figure}[!htbp]
    \centering
    \includegraphics[width=1.00\textwidth]{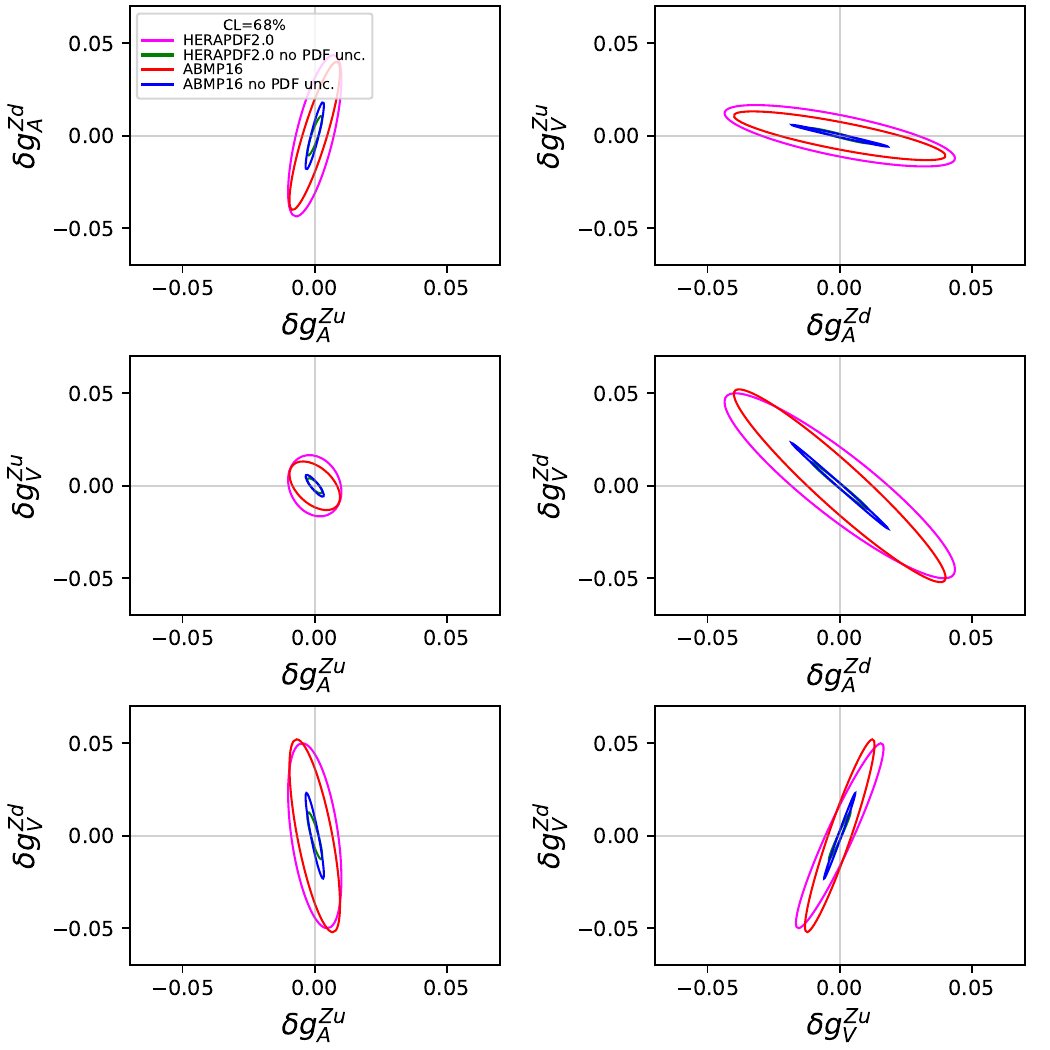}
    \caption{Same as in Fig.~\ref{fig:ell_pdf_unc} for the axial and vector couplings.}
    \label{fig:ell_pdf_unc_av}
\end{figure}

\begin{figure}[!htbp]
    \centering
    \includegraphics[width=0.65\textwidth]{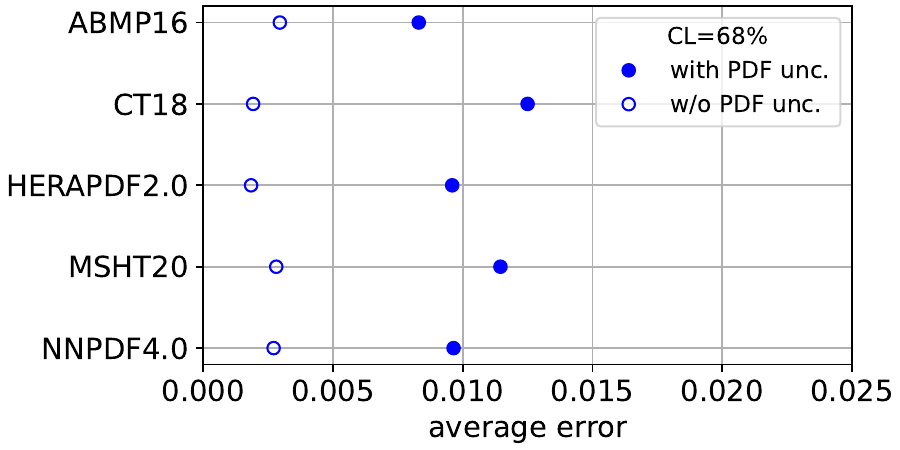}
    \caption{The average size of the uncertainties on the fitted corrections to the $Z$ couplings to $u$- and $d$-type quarks obtained using different PDF sets.}
    \label{fig:aver_pdf}
\end{figure}

\begin{figure}[!htbp]
    \centering
    \includegraphics[width=1.00\textwidth]{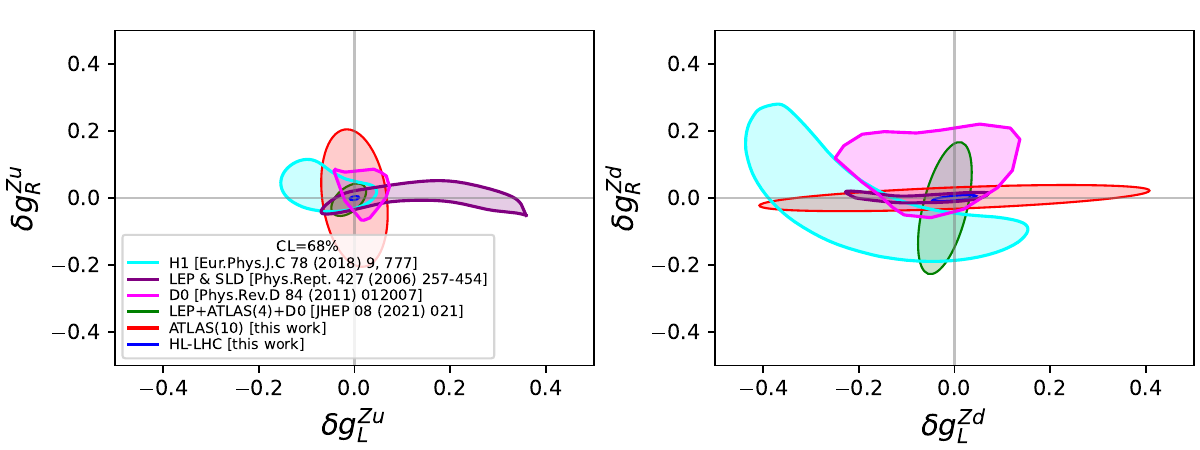}
    \caption{Allowed regions for all pairs of corrections to the $Z$ couplings to $u$- and $d$-type quarks obtained using HL-LHC pseudodata as well as different existing data sets.}
    \label{fig:ell_past}
\end{figure}

\begin{figure}[!htbp]
    \centering
    \includegraphics[width=1.00\textwidth]{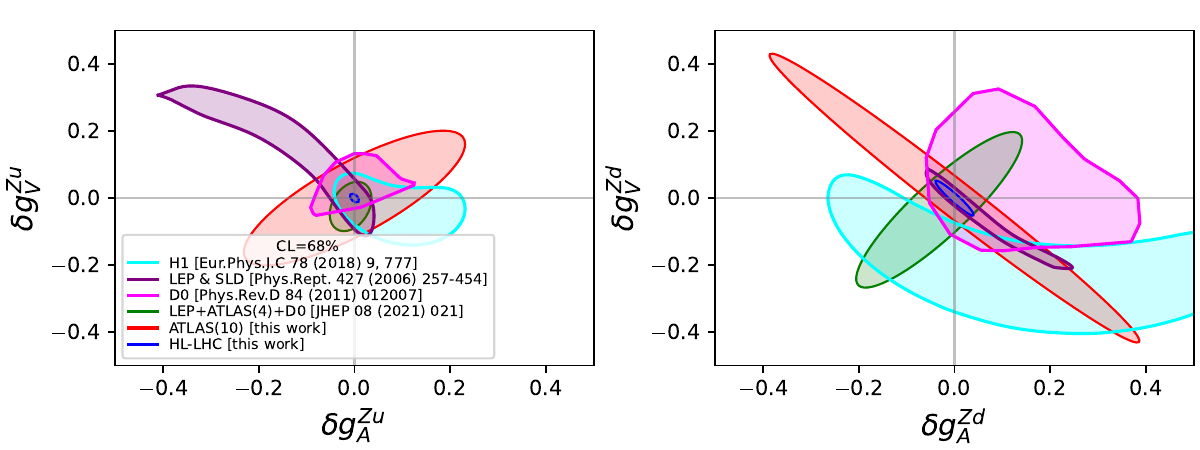}
    \caption{Same as in Fig.~\ref{fig:ell_past} for the axial and vector couplings.}
    \label{fig:ell_past_av}
\end{figure}

\begin{figure}[!htbp]
    \centering
    \includegraphics[width=1.00\textwidth]{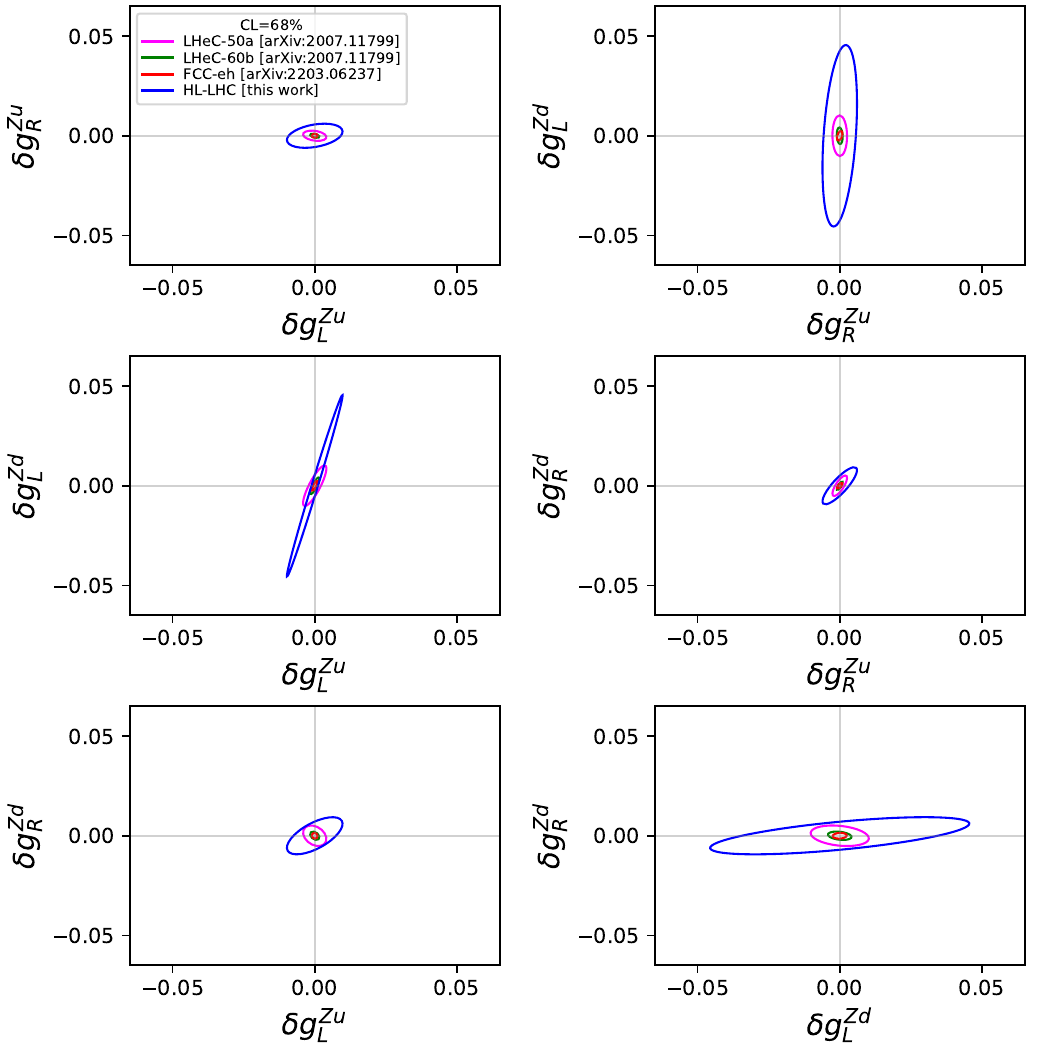}
    \caption{Allowed regions for all pairs of corrections to the $Z$ couplings to $u$- and $d$-type quarks obtained using HL-LHC pseudodata compared to the ones for different future experiments.}
    \label{fig:ell_future}
\end{figure}

\begin{figure}[!htbp]
    \centering
    \includegraphics[width=1.00\textwidth]{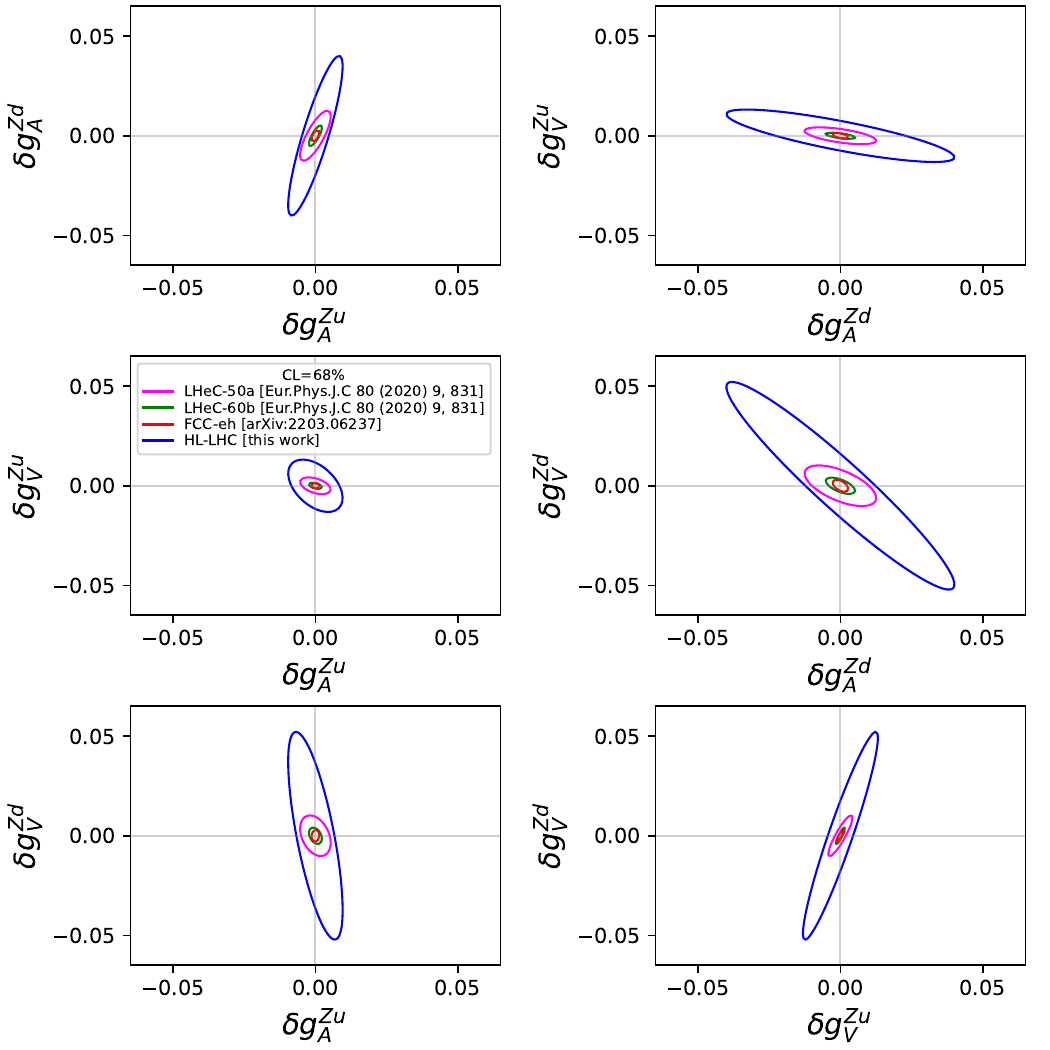}
    \caption{Same as in Fig.~\ref{fig:ell_future} for the axial and vector couplings.}
    \label{fig:ell_future_av}
\end{figure}

The results of the fit are presented in Figs.~\ref{fig:ell_pdf}--\ref{fig:ell_future_av} as allowed regions for different pairs of corrections to the $Z$ couplings to $u$- and $d$-type quarks at confidence level (CL) of 68\%.%
\new{Note that we fit four couplings at a time, while for presentation purposes we show 2D projection plots with different pairs of couplings.}
The resulting uncertainties on the couplings to the $d$-type quarks are roughly a factor of two larger than the corresponding uncertainties for the $u$-type quarks. 
A similar impact  of the \afb measurements on the PDFs was found in Ref.~\cite{Accomando:2019vqt}: it was shown that these measurements are most sensitive to the weighted sum $(2/3)u_v+(1/3)d_v$ of the valence $u$- and $d$-quarks, and we report the same finding in the present study.
This is related to the valence quark content of the proton and also to the fact that the $d\bar{d}$ initiated process gets more suppressed at high rapidity (where the sensitivity to the couplings is largest) in comparison to the $u\bar{u}$ initiated ones~\cite{Accomando:2018nig}. 
The exact details of this effect depend on the quark PDF behaviour at high values of the partonic momentum fraction $x$~\cite{Fiaschi:2022wgl,Fu:2023rrs,Hammou:2023heg,Alekhin:2023uqx,Ball:2022qtp}.
We find a strong correlation (up to 0.95) between the different couplings, as illustrated in Figs.~\ref{fig:ell_pdf}--\ref{fig:ell_future_av}, while the correlation coefficients between the couplings and the quark PDFs are moderate (about $0.5$). Furthermore, the latter exhibit significant variability across different values of $x$, reflecting the complex correlation of the different parton distributions in the original PDF sets.

In Figs.~\ref{fig:ell_pdf} and \ref{fig:ell_pdf_av} the allowed regions are shown as obtained using different PDF sets.\footnote{The PDF uncertainties of the CT18 set were rescaled to CL=68\%.} 
These results are presented using either couplings to right- and left-handed quarks or axial-vector couplings.
Both the size and the shape of the allowed regions (i.e.\ the correlations between the fitted parameters) are similar, independent of the PDF set.
To illustrate the impact of the PDF uncertainties on the results, in Figs.~\ref{fig:ell_pdf_unc} and \ref{fig:ell_pdf_unc_av} we show the allowed regions obtained with and without including the PDF uncertainties into the fit. 
Furthermore, in Fig.~\ref{fig:aver_pdf} we show the average size of the uncertainties on the fitted couplings as obtained using different PDF sets.
The impact of the PDF uncertainties is sizable. Namely, after including them into the fit the uncertainties on the couplings increase by a factor of $\sim 3$, i.e.\ the resulting uncertainties on the couplings are dominated by the PDF uncertainties. When using different PDF sets, the size of the average uncertainty does not change by more than a factor of $1.5$ indicating a reasonable consistency between the size of the PDF uncertainties for the modern PDF sets. However, it is worth mentioning also that the PDF uncertainties are constrained by the pseudodata when using the profiling technique. 
%On the other hand, if different PDF sets are employed to generate pseudodata and to fit the couplings, one does not find consistent results. 
The behaviour of quark and antiquark densities at large Bjorken-$x$ varies significantly between different PDF sets~\cite{Accomando:2018nig,Accomando:2017scx,Fiaschi:2022wgl} strongly affecting the theoretical predictions for the \afb in the high rapidity bins. However the discrepancies between different PDF sets will foreseeably reduce after the inclusion of additional experimental data covering the large-$x$ regions~\cite{Alekhin:2023uqx}.
The \afb observable has proved particularly sensitive to quark and anti-quark densities in the large Bjorken-$x$ region~\cite{Fiaschi:2022wgl,Fiaschi:2021okg,Accomando:2019vqt}, which can be accessed through measurements of the asymmetry at high rapidities~\cite{Fiaschi:2021okg,Accomando:2019vqt}.
In order to provide consistent results, in the future such analyses should comprise a simultaneous fit of the proton PDFs and the couplings.

In Figs.~\ref{fig:ell_past} and \ref{fig:ell_past_av} we compare our results obtained for the HL-LHC\footnote{We show the results obtained using the ABMP16 PDF set, since only this set provides symmetric PDF errors which are easier to include in the PDF profiling.} with the other analyses of existing data from LEP, Tevatron, HERA and LHC. The results are presented using either couplings to right- and left-handed quarks or axial and vector couplings. Namely, we compare with the analysis of the H1 Collaboration at HERA~\cite{H1:2018mkk}, the LEP+SLD combination~\cite{ALEPH:2005ab}, the analysis of D0 Collaboration at Tevatron~\cite{D0:2011baz} and the analysis of LEP, ATLAS and D0 data from Ref.~\cite{Breso-Pla:2021qoe}.%
\footnote{\neww{Note that on these plots we compare published results, which are obtained with different methodology in some cases. E.g.\ in our work and Refs.~\cite{Britzger:2020kgg,Britzger:2022abi} the assumption $\delta g_{R,L}^{Zd} = \delta g_{R,L}^{Zs} = \delta g_{R,L}^{Zb}$, $\delta g_{R,L}^{Zu} = \delta g_{R,L}^{Zc}$ was used, while in Ref.~\cite{Breso-Pla:2021qoe} the couplings to quarks of different flavours were fitted separately.
Also, Refs.~\cite{ALEPH:2005ab} and \cite{Breso-Pla:2021qoe} differ in the treatment of higher-order interference effects between the SM and new physics, universality assumption for down-type quarks, and the number of fitted parameters.}}
In addition to the HL-LHC results, we present our results of analyzing all available 10 bins from the ATLAS measurement of \afb~\cite{ATLAS:2018gqq}, while only $4$ bins at the $Z$ peak were used in the analysis of Ref.~\cite{Breso-Pla:2021qoe}. For the analysis of ATLAS data, we set the central data points to the theoretical predictions obtained at LO, while we use the data statistical and systematic uncertainties, as well as the PDF uncertainties.
\neww{The analysis of ATLAS data follows the procedure used for the HL-LHC pseudodata.}
Such a procedure provides credible uncertainties on the $Z$ couplings, while in order to get meaningful central values one would need to use theoretical calculations with higher-order QCD corrections. 
\neww{Given that the experimental uncertainties of the ATLAS data are larger than the uncertainties of the HL-LHC pseudodata, the PDF uncertainties play a moderate role in this case.}
The level of precision expected at the HL-LHC outperforms any existing data sets~\cite{ALEPH:2005ab,Breso-Pla:2021qoe,D0:2011baz,H1:2018mkk,ZEUS:2016vyd,Abt:2016zth}. \new{Also, we note that using all the $10$ data bins from the ATLAS measurement~\cite{ATLAS:2018gqq} provides more information on some of the couplings (e.g.\ $\delta g_R^{Zd}$) compared to the $4$ bins at the $Z$ peak together with the LEP and D0 data which were used in the analysis of Ref.~\cite{Breso-Pla:2021qoe}.}

\begin{figure}[!htbp]
    \centering
    \includegraphics[width=0.495\textwidth]{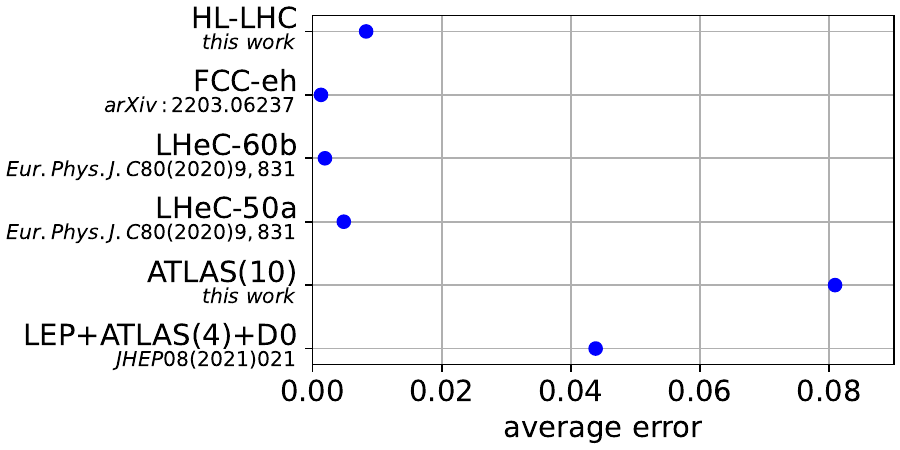}
    \includegraphics[width=0.495\textwidth]{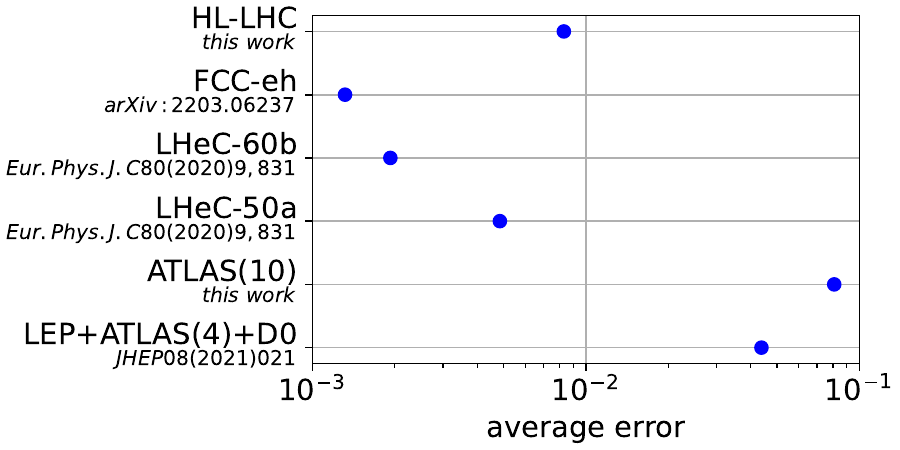}
    \caption{The average size of the uncertainties on the fitted corrections to the $Z$ couplings to $u$- and $d$-type quarks for different future experiments using the linear (left) or logarithmic (right) scale.}
    \label{fig:aver}
\end{figure}

In Figs.~\ref{fig:ell_future} and \ref{fig:ell_future_av} we compare the results obtained for the HL-LHC with the results expected at the future colliders currently under discussion,  LHeC~\cite{Britzger:2020kgg,LHeC:2020van} and FCC-eh~\cite{Britzger:2022abi,FCC:2018byv}. For the LHeC, two electron beam energies of $50$ or $60$ GeV are considered, and two assumptions on the uncertainties.
The results are provided for the so-called aggressive uncertainty scenario for $E_e = 60$ GeV, and the conservative one for $E_e = 50$ GeV (further details can be found in Ref.~\cite{Britzger:2020kgg}). Furthermore, in Fig.~\ref{fig:aver} the average size of the uncertainties which can be obtained using current and future data sets are compared. A sub-percent level of precision is expected at the LHeC, FCC-eh and HL-LHC, which is one order of magnitude better than what can be obtained using existing data sets from LEP, Tevatron, HERA and LHC. More precisely, the average size of the uncertainties which are expected at the FCC-eh are a factor of 6 better than the one from our HL-LHC expectation, while the LHeC uncertainties are only 1.7--4 times smaller (depending on the scenario) than our HL-LHC results.
Note that uncertainties may be also reduced at FCC-hh collider, due to increased cross section and luminosity compared to the HL-LHC. This could be studied using similar methods as those used for the HL-LHC in this paper. However, it requires a dedicated investigation of the detector acceptance and is beyond the scope of the current analysis.

\clearpage
\section{Conclusions}
\label{sec:conclusions}

We have studied the possibility to improve constraints on the $Z$ couplings to the $u$- and $d$-type quarks using the future measurements of \afb at the HL-LHC with $3000$ fb${}^{-1}$. We have investigated in detail the dependence of the \afb on the various couplings in different regions of the invariant mass of the lepton pair, and we have shown that a wide range of $45<M(ll)<145$ GeV is profitable to constrain the couplings. Furthermore, a measurement as function of the rapidity of the lepton pair provides a significant added value. Thus, we suggest double-differential measurements of the \afb as a function of both the invariant mass and rapidity of the lepton pairs\neww{, as done e.g.\ in Refs.\cite{ATLAS:2018gqq,Breso-Pla:2021qoe}}. Our quantitative analysis of the impact of the binning scheme on the precision of the extracted couplings suggests the choice of a specific binning scheme which ensures a substantial sensitivity to the couplings in such a measurement.

The resulting uncertainties on the couplings to the $d$-type quarks are found to be approximately a factor of two larger than the corresponding uncertainties for the $u$-type quarks.
Since the \afb observable is strongly sensitive to the proton PDFs, we find a significant dependence of the results on the PDF set used for such study, as was checked using the ABMP16, CT18, HERAPDF2.0, MSHT20 and NNPDF4.0 PDF sets. Preferably, in the future such analyses should comprise a simultaneous fit of the proton PDFs and the couplings.

The results were compared with the existing analyses of the LEP, HERA, Tevatron and LHC data, as well as with the results which are expected at the future colliders LHeC and FCC-eh. The uncertainties on the $Z$ couplings to the $u$- and $d$-type quarks at the HL-LHC are expected at percent level, thus outperforming by an order of magnitude any determinations of these couplings using existing data sets. This level of precision is similar, but a little inferior to the one which is expected at the LHeC and FCC-eh.

\acknowledgments

%csm put in alphabetical order
The work of S.~M. has been supported in part by the Bundesministerium f\"ur Bildung und Forschung under contract 05H21GUCCA.
The work of O.~Z. has been supported by the {\it Philipp Schwartz Initiative} of the Alexander von Humboldt foundation.
\neww{We thank Radja Boughezal and Yingsheng Huang for providing predictions for the SMEFT contributions of the four-fermion operators to \afb.}
\neww{We thank Adam Falkowski for the valuable comments on the comparison of the results from Refs.~\cite{ALEPH:2005ab} and \cite{Breso-Pla:2021qoe}.}

\neww{
\appendix
\section{Study of the impact of four-fermion SMEFT operators}
\label{sec:4f}

While the vertex corrections to the vector boson couplings to fermions scale as $M_Z^2/\Lambda^2$, the four-fermion interactions scale as $M_{ll}^2 /\Lambda^2$ and can be safely neglected on-peak, but potentially may become large at $M_{ll}>M_Z$. 
In Fig.~\ref{fig:4f} we compare the impact on \afb of the linear contributions of two four-fermion operators $C_{lq}^{(1)}$ and $C_{eu}$ from Ref.~\cite{Boughezal:2023nhe} with the impact of the vertex corrections to the couplings as a function of $M_{ll}$. 
For the four-fermion operators, the assumptions are $C=1$ and $\Lambda = 4$ TeV as typically used in the literature~\cite{Boughezal:2023nhe}, while for the vertex corrections we have varied the couplings by their uncertainties which were obtained in our study for the HL-LHC scenario with ABMP16 PDF set (see Fig.~\ref{fig:ell_pdf}).
As expected, the impact of the four-fermion operators is strongly suppressed at $M_{ll} \sim M_Z$, but it grows off-peak and becomes comparable to the impact of the vertex corrections to the couplings at $M_{ll} \approx 150$ GeV.
Since in our analysis we have chosen the region $M_{ll}<145$ GeV, we did not include the four-fermion operators.%
\footnote{For our sensitivity study even larger contributions from four-fermion operators at a percent level should not affect the main conclusions regarding the sensitivity to the SM couplings.}

\begin{figure}[!htbp]
    \centering
    \includegraphics[width=1.00\textwidth]{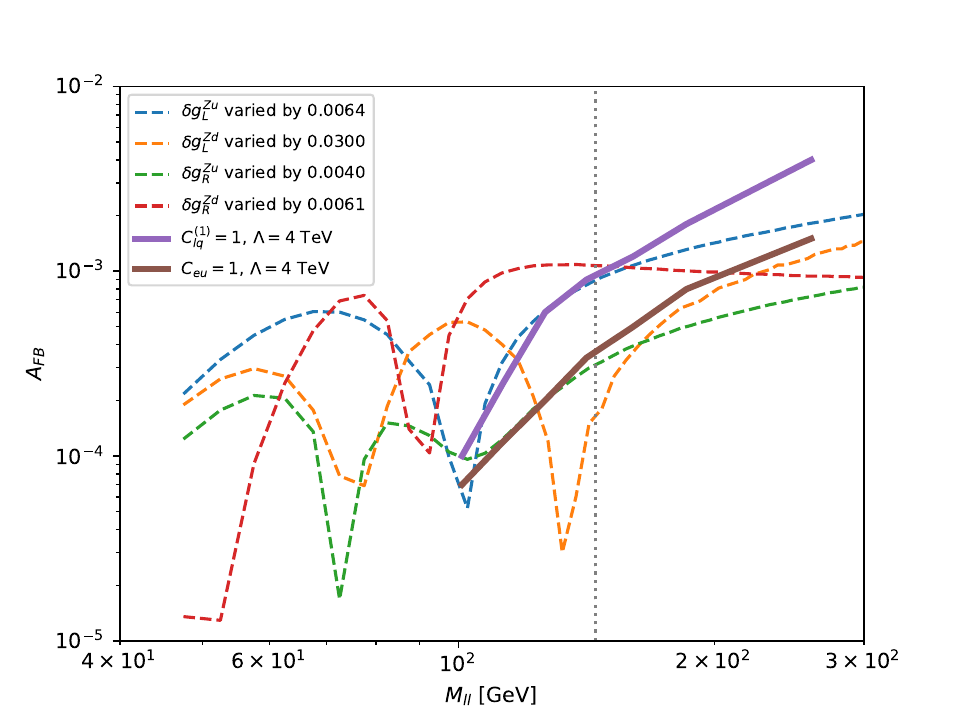}
    \caption{The linear contributions of four-fermion operators $C_{lq}^{(1)}$ and $C_{eu}$ and the contributions of the vertex corrections to couplings $\delta g_L^{Zu}$, $\delta g_L^{Zd}$, $\delta g_R^{Zu}$ and $\delta g_R^{Zd}$ to \afb as a function of $M_{ll}$. The dotted vertical line shows the value of the cut $M<145$ GeV used in our analysis.}
    \label{fig:4f}
\end{figure}

In order to study a possible dependence of our results on the upper boundary of $M_{\ell\ell}$, we repeated our analysis for the HL-LHC scenario using the ABMP16 PDF set with pseudodata restricted to the region $M_{\ell\ell}<115$~GeV only.
The fitted couplings are shown in Fig.~\ref{fig:ell_pdf_m115}, where the nominal results obtained with $M_{\ell\ell}<145$ GeV are shown as well. On average, the uncertainties on the couplings increase by $7\%$ if the cut $M_{\ell\ell}<115$ GeV is applied. We consider this as a robustness cross check of our results. We note that with future precise measurements of \afb in a wide kinematic region extending to $M_{\ell\ell}>150$ GeV it should be possible to provide constraints on both the SM couplings and the four-fermion SMEFT operators, and leave this for future studies.

\begin{figure}[!htbp]
    \centering
    \includegraphics[width=0.80\textwidth]{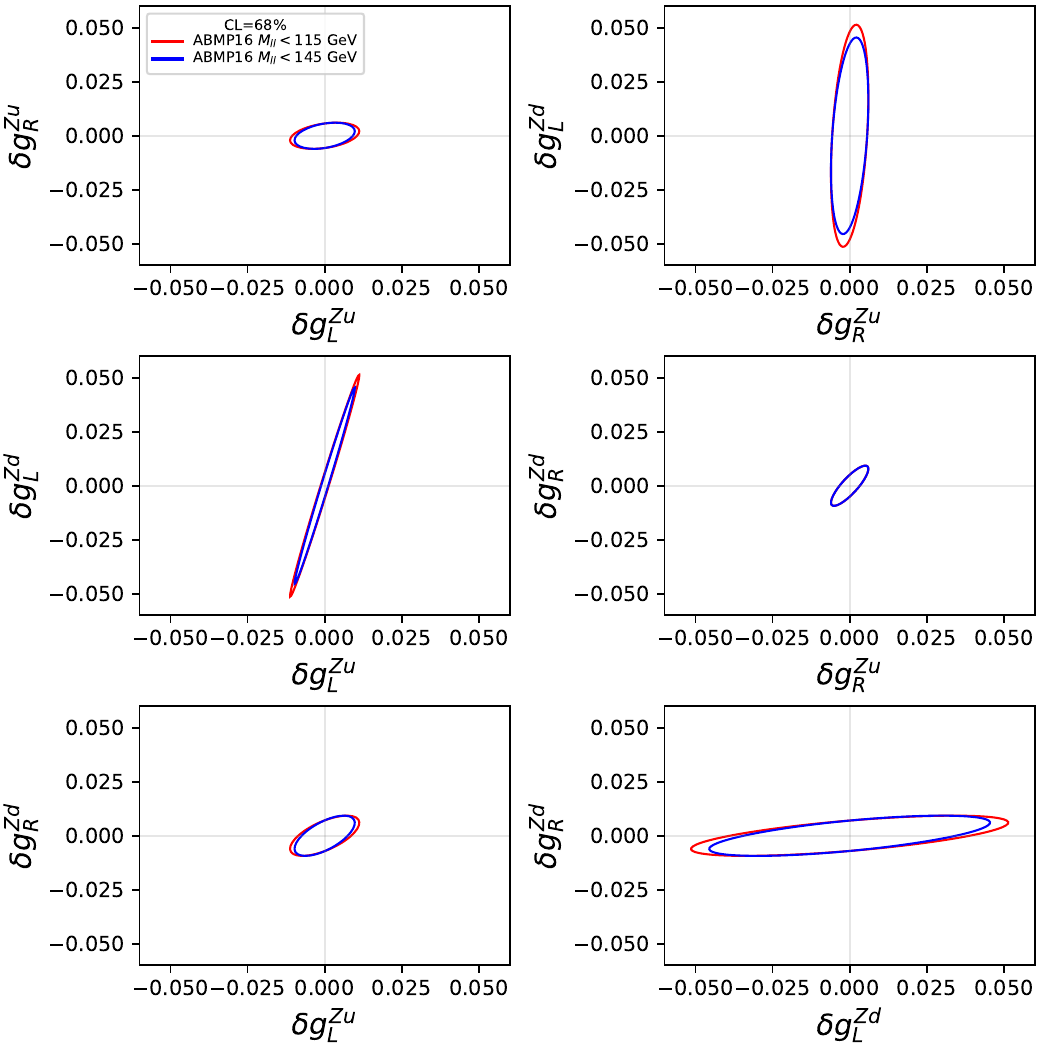}
    \caption{Allowed regions for all pairs of corrections to the $Z$ couplings to $u$- and $d$-type quarks obtained in the HL-LHC scenario with the ABMP16 PDF set using the $M_{ll}<145$ GeV and $M_{ll}<115$ GeV cuts.}
    \label{fig:ell_pdf_m115}
\end{figure}
}

\bibliographystyle{JHEP} 
\bibliography{afbsmeft}

\end{document}